\documentclass[10pt,english,journal]{IEEEtran}
\usepackage[T1]{fontenc}
\usepackage[latin9]{inputenc}
\usepackage{geometry}
\usepackage{amsmath}
\usepackage{xpatch}
\usepackage{amssymb}
\usepackage{esint}
\usepackage{mathtools}
\usepackage{amsthm}
\usepackage{nicefrac}
\usepackage{bigints}
\usepackage{mathtools}
\usepackage{blkarray, bigstrut}
\usepackage{physics}
\usepackage{calligra}
\usepackage{graphicx}
\usepackage{epstopdf}
\usepackage{dsfont}
\usepackage{array,ragged2e}
\usepackage{enumitem}
\usepackage{etoolbox}
\usepackage{babel}
\usepackage[nopar]{lipsum}
\usepackage{psfrag}
\usepackage[ruled,lined,linesnumbered]{algorithm2e}
\usepackage{algorithmic}
\usepackage{algorithm2e}
\usepackage{balance}
\usepackage{float}
\usepackage{hyperref}

\usepackage{subcaption}

\SetKw{KwBy}{by}

\makeatletter
\newcommand{\removelatexerror}{\let\@latex@error\@gobble}
\makeatother

\graphicspath{ {Figures/} }

\xpatchcmd{\proof}{\hskip\labelsep}{\hskip5\labelsep}{}{}  
\makeatletter
\xpatchcmd{\proof}{\@addpunct{.}}{\@addpunct{:}}{}{}
\makeatother

\renewcommand\[{\begin{equation}}
\renewcommand\]{\end{equation}} 
\usepackage{listings}
\usepackage{fancyvrb}
\usepackage{framed}

\usepackage{courier}
\usepackage[usenames,dvipsnames,table]{xcolor}

\definecolor{dkgreen}{rgb}{0,0.3,0}
\definecolor{gray}{rgb}{0.5,0.5,0.5}

\DeclarePairedDelimiter\floor{\lfloor}{\rfloor}

\makeatletter
\newcommand*{\rom}[1]{\expandafter\@slowromancap\romannumeral #1@}
\makeatother

\usepackage{siunitx}
\usepackage{tabu}
\usepackage{booktabs}
\usepackage{multirow}
\usepackage{capt-of}
\usepackage{array}
\usepackage{arydshln}
\DeclareMathOperator*{\argmax}{arg\,max}

\newcommand{\comment}[1]{}
\newcommand{\change}[1]{{\color{black} {#1}}}

\begin{document}

\title{
On the Specialization of FDRL Agents for Scalable and Distributed 6G RAN Slicing Orchestration 
}


\author{
Farhad~Rezazadeh,~\IEEEmembership{Student~Member,~IEEE}, Lanfranco~Zanzi,~\IEEEmembership{Member,~IEEE}, Francesco~Devoti,~\IEEEmembership{Member,~IEEE} Hatim~Chergui,~\IEEEmembership{Senior~Member,~IEEE}, Xavier~Costa-P\'erez,~\IEEEmembership{Senior~Member,~IEEE}, and~Christos~Verikoukis,~\IEEEmembership{Senior~Member,~IEEE}

\IEEEcompsocitemizethanks{
Copyright (c) 2015 IEEE. Personal use of 
this material is permitted. However, permission to use this material for any other purposes must be 
obtained from the IEEE by sending a request to pubs-permissions@ieee.org.}
\IEEEcompsocitemizethanks{This work was partially funded by the Spanish Government (initially by MICCIN and since November 2021 by the Next Generation EU program) under Grant PCI2020-112049 and by the Electronic Components and Systems for European Leadership Joint Undertaking (JU) under grant agreement No. 876868. This JU receives support from the EU`s H2020 research and innovation programme and Germany, Slovakia, Netherlands, Spain, Italy, and in part by the EU H2020 projects MonB5G (871780), 5GMediaHUB (101016714), and OPTIMIST
(872866).
}
\IEEEcompsocitemizethanks{\IEEEcompsocthanksitem F. Rezazadeh is with the Telecommunications Technological Center of Catalonia (CTTC) and  Technical University of Catalonia (UPC), Barcelona, Spain (e-mail: frezazadeh@cttc.es).}
\IEEEcompsocitemizethanks{\IEEEcompsocthanksitem L. Zanzi and F. Devoti are with the NEC Laboratories Europe, Heidelberg, Germany (e-mails: lanfranco.zanzi@neclab.eu, francesco.devoti@neclab.eu).}
\IEEEcompsocitemizethanks{\IEEEcompsocthanksitem H. Chergui is with the Telecommunications Technological Center of Catalonia (CTTC), Barcelona, Spain (e-mail: hchergui@cttc.es).}
\IEEEcompsocitemizethanks{\IEEEcompsocthanksitem Xavier Costa-P\'erez is with NEC Laboratories Europe, i2CAT Foundation, and ICREA, Barcelona, Spain (e-mail: xavier.costa@neclab.eu).}
\IEEEcompsocitemizethanks{\IEEEcompsocthanksitem C. Verikoukis is with the University of Patras, ATHENA/ISI, and IQUADRAT Informatica, Barcelona, Spain (e-mail: cveri@ceid.upatras.gr).}
}

\maketitle

\begin{abstract}
Network slicing enables multiple virtual networks to be instantiated and customized to meet heterogeneous use case requirements over 5G and beyond network deployments. However, most of the solutions available today face scalability issues when considering many slices, due to centralized controllers requiring a holistic view of the resource availability and consumption over different networking domains. In order to tackle this challenge, we design a \emph{hierarchical architecture to manage network slices resources in a federated manner}. 
Driven by the rapid evolution of deep reinforcement learning (DRL) schemes and the Open RAN (O-RAN) paradigm, we propose a set of traffic-aware local decision agents (DAs) dynamically placed in the radio access network (RAN). These federated decision entities tailor their resource allocation policy according to the long-term dynamics of the underlying traffic, defining \emph{specialized} clusters that enable faster training and communication overhead reduction.
Indeed, aided by a traffic-aware agent selection algorithm, our proposed \emph{Federated DRL} approach provides higher resource efficiency than benchmark solutions by quickly reacting to end-user mobility patterns and reducing costly interactions with centralized controllers.
\end{abstract}


\section{Introduction}

\begin{figure}[t!]
\centering
\includegraphics[clip, trim = 6cm 5.5cm 13cm 3cm, width=0.85\columnwidth]{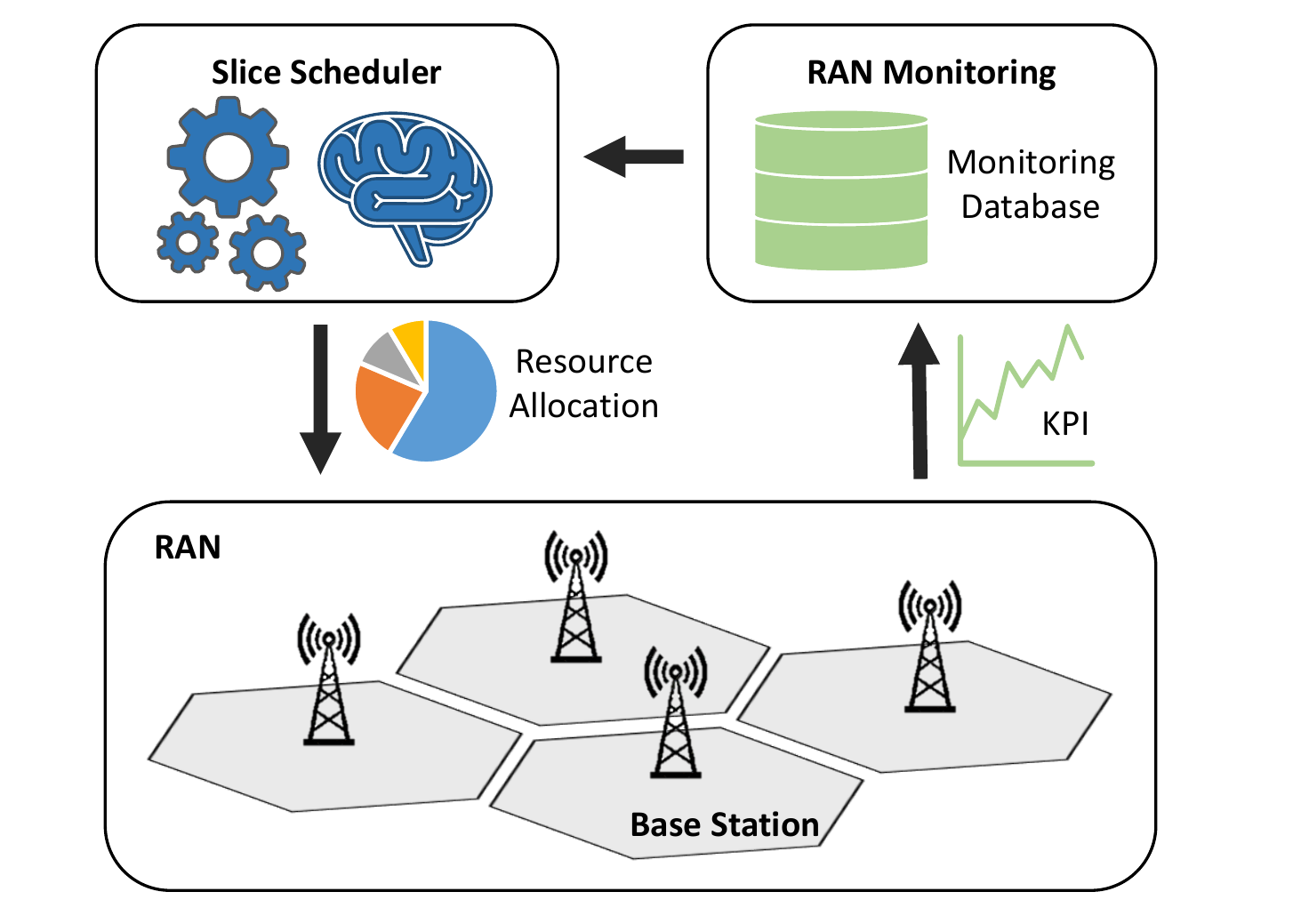}
\caption{ \small RAN resource allocation in network slicing.}
\label{fig:problem}
\vspace{-0.6cm}
\end{figure}

 \IEEEPARstart{V}{ehicle-to-everything} \change{(V2X) communication, Internet of things (IoT), augmented/virtual reality (AR/VR), are just some examples of emerging use-cases in 5G/6G verticals that need to co-exist over a common physical infrastructure.
However, the highly heterogeneous performance requirements in terms of bandwidth, latency, and reliability, exacerbate the need for orchestration solutions able to accommodate such services in a resource and cost-efficient manner.}
Network slicing represents a promising technology able to address such a challenging scenario, by enabling the setup of multiple logical and virtualized network instances, namely \emph{slices}, on top of a common physical mobile network infrastructure~\cite{TVT_2}. Given the cloud nature of these resources, the networking resources associated to each slice can be dynamically orchestrated and tailored to meet the performance requirements of running services.
In this context, temporal variations of the traffic demand deeply complicate resource planning and allocation tasks, especially in the radio access network (RAN) domain where resource allocation \cite{Zhou_sli} decisions, e.g., in terms of bandwidth, must cope with the additional variability inherent of the wireless channel and end-user's mobility.
Traditional RAN slicing solutions envision a centralized controller with a holistic and real-time view of the network, especially about resource utilization, availability, and real-time wireless channel statistics, as depicted in Fig.~\ref{fig:problem}. However, similar approaches suffer from scalability issues in real deployments, where the amount of monitoring information to be exchanged, together with the \change{large} number of base stations (BSs), make it practically impossible to devise optimal resource allocation schemes in a timely and resource-efficient manner~\cite{f-centralized}. 
Motivated by the need for a more cost-effective and agile RAN, the Open RAN (O-RAN) Alliance recently presented a vendor-neutral alternative way of building mobile networks~\cite{ORAN_White_paper}, based on disaggregated hardware and interoperable interfaces that allow secure network sharing by means of virtualization. 
\change{Despite the revolutionary approach, it is still not clear how to efficiently support slicing scenarios~\cite{masssive-slicing} characterized by a large number of vertical services.} Therefore, we take on this challenge and propose a hierarchical architecture for network slice resource orchestration. In particular, given the variable spatio-temporal distribution of mobile traffic demands~\cite{AztecInfocom20}, we envision the dynamic setup of a network of local decision agents (DAs) as virtual software instances co-located within the Near-Real Time RAN Intelligent Controller (Near-RT RIC) premises, able to access local RAN monitoring information and extract local knowledge without the need of a centralized entity performing decisions on aggregated information. Our framework leverages a dynamic agent selection mechanism based on local traffic conditions similarity, which enables more efficient information exchange and collaboration among groups of local DAs, while specializing their decision policy.
The benefit coming from our approach are several: $i$) it enables resource allocation at the edge of the network, thus accounting for more timely and accurate information, $ii$) the amount of control information that needs to cross the network to reach the central controller dramatically decreases, thus reducing overhead towards the core network, $iii$) by allowing information exchange among local DAs, we enable the provisioning of federated learning schemes to further enrich the capabilities of the DAs.
In fact, DAs will not only learn from a local observation space, but also leverage information coming from other (statistically different) RAN nodes, thus improving the generalization of the learning procedure.

The main contributions of our paper can be summarized as follows:
\begin{itemize}
    \item We cast the RAN resource allocation problem as an optimization problem, focusing on minimizing the traffic exceeding the service level agreement (SLA) and assessing its complexity.
    \item We propose a distributed architecture for RAN slice resource orchestration based on deep reinforcement learning (DRL), composed of multiple artificial intelligence (AI)-enabled decision agents that perform local radio allocation decisions without the need for a centralized control entity.
    \item We design a federated learning (FL) scheme composed of multiple parallel layers, one for each slice, to enhance the capabilities of the local decision-making process,  following the recent development of the Open RAN architecture.
    \item We further improve the decision process by dynamically defining the subset of decision agents to be involved in the federation process, based on long-term slice traffic demands variations and their temporal similarities.
    \item We validate our hierarchical architecture and assess its capabilities in realistic scenarios by means of an exhaustive simulation campaign, accounting for a wide geographical area and thousands of end-users.
\end{itemize}

The remainder of this paper is as follows:
Sec.~\ref{sec:related} provides an overview of the related works in the field.
Sec.~\ref{sec:scenario} formulates our problem and describes the considered scenario.
Sec.~\ref{sec:architecture} presents the main building blocks of our solution, describing the interaction among the different entities.
Sec.~\ref{sec:ORAN} highlights the compliance of our solution with respect to the O-RAN architecture.
Sec.~\ref{sec:perf_eval} validates the design principles of our solution through a comprehensive simulation campaign.
Finally, Sec.~\ref{sec:conclusion} provides the final remarks and concludes this paper.

\section{Related Work}
\label{sec:related}

AI-driven approaches applied to mobile networks have recently gained momentum in distributed resource control and management tasks. In this context, DRL \cite{mano-far2s,ans2-far} and federated DRL (FDRL) \cite{Zhang_slice} stand out among a multitude of different approaches and are at the center of a strong research interest, especially in the field of automated resource orchestration. 
\change{ The authors of \cite{rec1} consider the sum power minimization problem based on jointly optimizing resource allocation, user association, and power control in a multi-access edge computing (MEC) system. In this intent, they propose a multi-agent federated reinforcement learning algorithm to solve centralized method limitations and privacy concerns. The simulation results shown that the proposed approach provides lower maximal latency, lower maximal computation capacity,
higher CPU cycles for the tasks, and higher data rate. Due to enhancing spectrum utilization in new generation wireless communication technologies, the authors in \cite{rec2} invoke an FDRL approach to accelerate learning convergence in edge nodes. 
In \cite{rec3}, the authors investigate the decentralized joint optimization of channel selection and power control for V2V communication, proposing a federated multi-agent DRL (Fed-MARL) approach to satisfy the reliability and latency requirements of V2V communication, and maximize the transmit rates of cellular links. The results have shown how the federation of local DRL models coming from different V2V agents can tackle the limitations of partial observability of the entire network, resulting in superior perfomances over baseline approaches in terms of communication rate and packet delivery rate. 
Targeting at implementing the open RAN (O-RAN) with virtualized network components, the authors of~\cite{rec6} proposed an FDRL-based with a global model server installed in the intelligent controller (RIC) to update the deep Q-networks parameters. This approach can mitigate load balancing and frequent handovers in the massive base station deployment. Following O-RAN standardization, the numerical results have demonstrated the proposed method enables UE to effectively maximizes the long-term throughput and avoids frequent handovers.}
The authors of~\cite{FMA_HUY} develop a DRL algorithm for the resource allocation in a mobility-aware FL network, optimizing the number of successful transmissions while minimizing energy and channel costs. In~\cite{Fdrl_seif}, the authors propose a federated network slicing scheme based on DRL techniques for channels and bandwidth allocation in the context of industrial IoT (IIoT), highlighting significant performance improvements when compared against centralized strategies.
In~\cite{DeepRL_Infocom} the authors model the network utility maximization problem and exploit DRL techniques such as deep Q-learning to solve the decision-making task. Their work highlights significant improvements over key performance indicators (KPIs) and networking metrics, such as throughput and latency. This solution however exploits a centralized approach that aims to solve a global optimization, therefore limiting the individual network slices in the management of their own resources.
A similar problem has been addressed by~\cite{EdgeSlice}, which however proposes a decentralized resource orchestration system to automate dynamic end-to-end network slicing resource management in wireless edge computing networks. The proposed architecture makes use of a central performance coordinator entity and multiple orchestration agents. This work however provides limited details on inter-agent information exchange aspects.
Also~\cite{Scalable_Orchestration} proposes a DRL approach for the orchestration of service function chains in NFV-enabled networks, addressing both placement-error-rate-based and reward-based federated weighed strategies, showing significant convergence performance, higher average reward, and smaller average resource consumption in a variety of networking scenarios.

From the viewpoint of real-time inter-slice resource management and yield an intelligent strategy, the authors of~\cite{TVT_5} design a graph attention multi-agent Reinforcement Learning to cope with frequent BS handover. The simulation results have demonstrated that the proposed approach is effective to enhance the cooperation for the multi-BS system in RAN while satisfying the strict SLA requirements. In~\cite{TVT_4}, the authors propose a multi-agent reinforcement learning approach for RAN capacity sharing, showcasing better scalability and faster learning in comparison to single-agent approaches. More recently, the authors of~\cite{NODL} develop an FL framework in the context of fog computing, focusing on the distribution of training tasks. The numerical results show that the proposed network-aware scheme significantly improves network resource utilization while achieving comparable accuracy.
%

Following this state-of-the-art overview, the key novelty of our approach relies on the exploitation of distributed RAN information to design a new class of specialized agents that collaborate in homogeneous clusters via a federation layer,  which leads to scalable and stable decision under highly dynamic traffic conditions and then proposed framework is also mapped to O-RAN. To the best of our knowledge, this is the first work to propose an FDRL framework in the context of distributed radio resource management, by adopting dynamic agent selection to improve specialization of agents and reduce communication overhead.

\begin{table}[!t]
\caption{\change{Notation Table}}
\label{tab:model_var_and_par}
\centering
\begin{tabular}{@{}lc@{}}\toprule

\textbf{Notation} & \textbf{Description} \\ \midrule
$\mathcal{B}$ & Set of BSs\\ \hdashline
$\mathcal{I}$ & Set of slices\\ \hdashline
$C_b$ & Capacity of BS $b\in\mathcal{B}$\\ \hdashline
$\Lambda_i$ & Latency requirement of slice $i\in\mathcal{I}$\\ \hdashline
$\lambda_i$ & Throughput requirement of slice $i\in\mathcal{I}$\\ \hdashline
$\mathcal{T}$ & Set of decision intervals\\ \hdashline
$\epsilon$ & Duration of decision intervals $t\in\mathcal{T}$\\ \hdashline
$a_{i,b}^{(t)}$ & PRB allocation for slice $i\in\mathcal{I}$ at BS $b\in\mathcal{B}$\\
 & taken at time interval $t\in\mathcal{T}$\\ \hdashline
$\sigma_{i,b}^{(t)}$ & Average SNR experienced by the users of\\
& slice $i\in\mathcal{I}$ at BS $b\in\mathcal{B}$ in time interval $t\in\mathcal{T}$\\ \hdashline
$\varphi_{i,b}^{(t)}$ & Instantaneous traffic demand of the users of\\
& slice $i\in\mathcal{I}$ at BS $b\in\mathcal{B}$ in time interval $t\in\mathcal{T}$\\ \hdashline
$d_{i,b}^{(t)}$ & Traffic of slice $i\in\mathcal{I}$ at BS $b\in\mathcal{B}$\\
& dropped in time interval $t\in\mathcal{T}$\\ \hdashline
$\iota$ & Minimum PRB allocation\\ 
\hdashline

\change{$E_{i,b}^{(t)}$} & \change{Expected transmission latency}\\ 
\hdashline
\change{$\nu_i^{(t)}$} & \change{Amount of available capacity left by}\\& \change{previous decisions of other agents} \\ 
\hdashline
\change{$\alpha_{i}^{(t)}$} & \change{Allocation gap}\\ 
\hdashline

\change{$\rho^{(t)}_{\text{up}}$} & \change{Upper boundary of allocation gap}\\ 
\hdashline

\change{$\rho^{(t)}_{\text{lower}}$} & \change{Lower boundary of allocation gap}\\ 
\hdashline
\change{$r_i^{(t)}$} & \change{Instantaneous reward of the $i$-th agent}\\ 
\hdashline

\change{$P_i^{(t)}$} & \change{Penalty of $i$-th agent}\\ 
\hdashline

\change{$\eta_i$} & \change{Penalty coefficient}\\ 
\hdashline

\change{$Q_{\pi}$} & \change{Action-value function under a given policy $\pi$}\\ 
\hdashline

\change{$\gamma$} & \change{Discount factor}\\ 
\hdashline
\change{$\xi$} & \change{Learning rate}\\ 
\hdashline

\change{$\beta_{i}$} & \change{Experience buffer}\\ 
\hdashline

\change{$\theta_{i}^{(t)}$} & \change{Online network parameter}\\ 
\hdashline

\change{$\tilde{\theta}_{i}^{(t)}$} & \change{Target network parameter}\\ 
\hdashline

\change{$\Omega_{i}^{(t+1)}$} & \change{Global updated model}\\ 
\hdashline

\change{$\Psi$} & \change{Set of clusters}\\ 

\bottomrule
\end{tabular}
\end{table}

\section{Framework Overview}
\label{sec:scenario}

Our solution builds on the concept of slicing-enabled mobile networks~\cite{net-Slicing}, wherein multiple network tenants are sharing a portion, namely a \emph{slice}, of a common mobile network infrastructure, each one with predefined and dedicated networking resources to satisfy an SLA. Within the context of our paper, we focus on the RAN domain and consider the SLAs to be expressed in terms of maximum slice throughput and transmission latency. We define transmission latency as the average time the traffic belonging to a certain slice needs to wait within the base station transmission buffers before being served due to inter-slice scheduling procedures.
In the following, we overview the main system building blocks and model assumptions, finally introducing the mathematical formulation of the RAN resource allocation problem. We summarize parameters and variables describing the system model in Table~\ref{tab:model_var_and_par}.

Let us introduce a mobile network infrastructure composed of a set $\mathcal{B}$ of base stations (BSs), wherein a set of slices $\mathcal{I}$ is deployed. Each BS $b \in \mathcal{B}$ is characterized by a capacity $C_b$, expressed in terms of a discrete number of physical resource blocks (PRBs) of a fixed bandwidth. 
This resource availability must be divided into subsets of PRBs, and dynamically assigned to each network slice according to their real-time traffic demand and SLA requirements. 
\change{As part of the SLA between the network operator and the slice owner, we assume each network slice to come with predefined latency and throughput requirements defined by the variables $\Lambda_i$ and $\lambda_i$, respectively. In the context of our work, we focus on the RAN domain, and therefore consider as latency the queueing delay time experienced by the traffic while flowing through the scheduling processes of each base station.}
Let us consider a time-slotted system where time is divided into \emph{decision intervals} $t \in \mathcal{T} = \{ 1,2,\dots,T\}$.
The PRB allocation decisions can be taken only at the beginning of each decision interval, whose duration $\epsilon$ may be decided according to the infrastructure provider policies, ranging from few seconds up to several minutes.

We assume the presence of a preliminary admission and control mechanism, e.g., the one presented in~\cite{RL-NSB}, to verify the admissibility of the current network slice setup within the available networking capacity, and focus our effort on meeting the resource allocation for the downlink traffic.
We envision the allocation of radio resources towards the end-users as a two-step process~\cite{foukas_orion}. Initially, once network slices are admitted into the system, the infrastructure provider schedules the assignment of slots of radio resources for each of the tenants. Then, based on the slice resource availability, each tenant may decide to enforce proprietary scheduling solutions towards its end-users, depending on use-case or business requirements~\cite{net-Slicing}. 
\change{Given the plethora of user to base station association and scheduling algorithms addressing the end-user resource allocation task~\cite{Spatial_Loads}, we do not address the intra-slice scheduling issue, but rather focus on the correct and fair dimensioning of the inter-slice PRB allocation.}
To this aim, we denote with the variable $a_{i,b}^{(t)}$ the PRB allocation decision for the $i$-th slice under the $b$-th BS taken at $t$-th decision time interval, and with $\sigma_{i,b}^{(t)}$ the signal-to-noise ratio (SNR) value expressing the wireless channel quality, averaged over the duration of a decision time interval $\epsilon$, and over the end-users of the $i$-th slice attached to the $b$-th BS. Similarly, we introduce $\varphi_{i,b}^{(t)}$ as the aggregated downlink traffic demand generated by the users of the $i$-th slice under the coverage area of the $b$-th BS within the $t$-th time interval. 
All together, we can formalize our problem as: \\
\noindent \textbf{Problem}~\texttt{RAN Resource Allocation}:
\label{prob:RAN_Slicing}
\begin{flalign}
\text{min} &  \lim\limits_{{T\rightarrow\infty}} \sum\limits_{t = 1}^T \mathbb{E}\left [\sum\limits_{i\in\mathcal{I}} d_{i,b}^{(t)} \right ] \hspace{-3cm}& &\\
\noindent\text{subject to:} \hspace{-1cm}& & &\nonumber\\
& E_{i,b}^{(t)}  \leq \Lambda_i, \hspace{-0.7cm}& \forall t \in \mathcal{T}, \forall i\in\mathcal{I}, \forall b\in\mathcal{B};&\\
& \sum\limits_{i\in\mathcal{I}} a_{i,b}^{(t)} \leq C_b, &\forall t \in \mathcal{T}, \forall b \in \mathcal{B};&\\
& a_{i,b}^{(t)}\in\mathbb{Z}_+, d_{i,b}^{(t)}\in\mathbb{R}_+, &\forall t \in \mathcal{T}, \forall i\in\mathcal{I},\forall b\in\mathcal{B};&
\end{flalign}
where $E_{i,b}^{(t)} = \mathbb{E}\left [ \frac{\varphi_{i,b}^{(t)}}{\Gamma \left(a_{i,b}^{(t)},\sigma_{i,b}^{(t)} \right)+d_{i,b}^{(t)}}\right ]$
defines the expected transmission latency, and $\Gamma(a, \sigma)$ is a function that translates the PRB allocation $a$ in the equivalent transmission capacity, given the experienced channel quality $\sigma$. 
The traffic demand generated within a decision interval might not be fully satisfied due to erroneous PRB allocation estimations, incurring in additional transmission \change{latency} due to traffic queuing at the base station. Therefore, we introduce 
the variable $d_{i,b}^{(t)}$ as a deficit value indicating the volume of traffic not served within the agreed slice latency tolerance $\Lambda_i$, and that is therefore dropped.
\change{
Due to fast traffic variations, slice resource allocation \cite{melike_sli} decisions at the RAN domain should be taken in a dynamic, proactive, and flexible way to avoid service and performance degradation. While advanced admission and control mechanisms could select the set of slices to be admitted to the system, and provide static resource allocation boundaries to satisfy the available capacity, the dynamic nature of the slice's traffic load and wireless channel statistics may lead to suboptimal performances.

Additionally, the optimization problem underlying RAN resource allocation, that is, fitting the requests of the slices maximizing the overall utilization by considering the limited resource availability of a BS, has been proven to be NP-Hard~\cite{RL-NSB}. 
In fact, this problem can be easily mapped into a knapsack problem instance, wherein the sum of allocated resources is bounded by the capacity of the radio interface, and the experienced latency, i.e., the cost, is minimized. This family of problems is well-known to be NP-Hard~\cite{Knapsack_NP_Complete}, resulting in a time complexity of $O(IC_b)$ in our scenario, where $I$ is the cardinality of the set $\mathcal{I}$, and $C_b$ is the base station resource availability in number of PRBs. 
In order to obtain a solution for the overall RAN deployment, the same problem should be solved for all the nodes in the network, therefore introducing scalability issues. 
Moreover, the centralization of all the necessary up-to-date monitoring information further exacerbates the complexity of this problem, which becomes impractical in real mobile networks characterized by thousands of RAN nodes~\cite{pi_ROAD}.
}



\section{A multi-agent architecture for RAN resource allocation}
\label{sec:architecture}

\begin{figure}[t!]
\centering
\includegraphics[clip,width=\columnwidth]{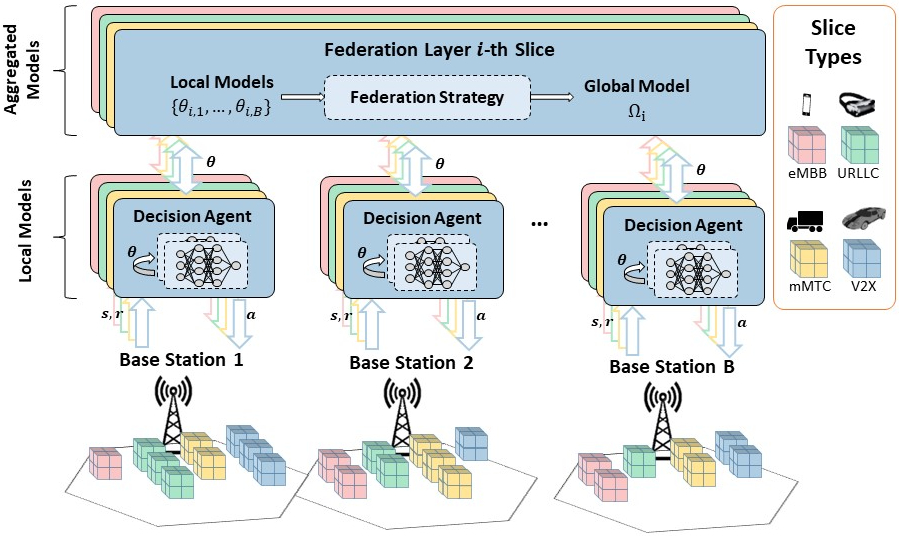}
\caption{\small Generic Federated DRL architecture for RAN slicing.}
\vspace{-5mm}
\label{fig:architecure}
\end{figure}

\change{In this paper,} we advocate for the adoption of an FDRL-based architecture to address the RAN slicing scenario. In particular, we rely on local DAs running as software instances within the premises of each BS, as shown in Fig.~\ref{fig:architecure}. Each agent is in charge of performing slice PRB allocation decisions based on local monitoring information coming from the underlying network monitoring system, or BS \emph{context}. We provide the details of our local decision algorithm later in Sec.~\ref{sec:DQN}.
Nevertheless, the distributed nature of RAN deployments, as well as the varying spatio-temporal behavior of mobile traffic traces~\cite{FurnoTMC2017}, make it difficult for an agent trained exclusively on complex and multi-variate monitoring metrics to address unknown statistical distributions of its base station context. 

To concurrently address the above issues, we introduce an FL layer that allows inter-agent information exchange, and expedites the learning procedure  local knowledge sharing. We provide the details of our FL approach in Sec.~\ref{sec:federated}.

\subsection{Local RAN Slicing via DDQN Agent}
\label{sec:DQN}

Deep Q-network (DQN) is a popular reinforcement learning~\cite{TVT_1} algorithm that evolves from the well-known concepts of Q-learning and neural network function approximation.
DQN represents a model-free approach. It stores the trajectory of experiences for each interaction with the environment in a replay buffer, as to update the network parameters without prior knowledge of the underlying environment statistics.
In the following, we will use the $i$ index interchangeably while referring to slices and DAs, assuming a one-to-one mapping of each DA with the corresponding network slice.
With focus on a single BS and a single decision interval 
the design choices of our DQN model are as follows:\\
\textbf{State Space $\mathcal{S}$} We define the state of the $i$-th agent associated to the $b$-th BS as a tuple of local monitoring information $s_i^{(t)} = \{ ( \sigma_i^{(t)}, \lambda_i^{(t)}, \nu_i^{(t)} ) \mid \forall i \in \mathcal{I} \},$ where $\sigma_i^{(t)}$ is the SNR value, averaged over the duration of a decision time interval experienced by the users of the $i$-th slice, $\lambda_i^{(t)}$ is the aggregated traffic volume generated by the $i$-th slice over the time decision duration $\epsilon$,
and $\nu_i^{(t)}$ is the amount of available capacity left by the previous decisions of other agents.
\\%
\textbf{Action Space $\mathcal{A}$} Without loss of generality, we define $\iota$ as the minimum PRB allocation step, or \emph{chunk size}, and assume that the PRB allocation decision of the $i$-th agent can only take values that are an integer multiple of $\iota$.
It results that $\mathcal{A} = \{ \iota \cdot k \mid k = \{0,1, \dots, \floor{\frac{C}{\iota}} \} \}$.
Such discrete action space allows controlling the dimensionality of the action space and positively influences the learning process~\cite{LACO}.
\\%
\textbf{Reward $\mathcal{R}$} We adopt an iterative reward-penalty approach to guide the agent learning procedure, which translates into maximizing a reward function. An accurate PRB allocation should concurrently guarantee the satisfaction of transmission latency $\Lambda_i$ and the traffic requirements $ \lambda_i^{(t)}$, while avoiding both under-provisioning and over-provisioning of resources. Given the instantaneous slice traffic volume $\varphi_i^{(t)}$, and the corresponding allocation decision $a_{i}^{(t)} \in \mathcal{A}$, we can identify an \emph{allocation gap} $\alpha_{i}^{(t)} = \Gamma(a_{i}^{(t)},\sigma_i^{(t)}) - \varphi_i^{(t)}$. To measure the goodness of the action, we therefore introduce two variables, namely $\rho^{(t)}_{\text{up}}$ and $\rho^{(t)}_{\text{lower}}$, which characterize the upper and lower boundaries of the allocation gap as $\rho^{(t)}_{\text{up}} = 2\cdot\Gamma(\iota^{(t)},\sigma_i^{(t)}) $ and $\rho^{(t)}_{\text{lower}} = -\Gamma(\iota^{(t)},\sigma_i^{(t)})$. Accordingly, we define the instantaneous reward $r^{(t)}_i \in \mathcal{R}$ of the $i$-th agent as: 
\[
    r_i^{(t)} =\begin{cases}
    \alpha^{(t)}_{i} -4 \rho^{(t)}_{\text{lower}}  & \text{if} \qquad \alpha^{(t)}_{i} < \rho^{(t)}_{\text{lower}}, \\
    (1- \frac{\alpha^{(t)}_{i}}{\rho^{(t)}_{\text{up}}}) \frac{\alpha^{(t)}_{i}}{\rho^{(t)}_{\text{up}}}  & \text{if} \qquad \rho^{(t)}_{\text{lower}} \leq \alpha^{(t)}_{i} \leq \rho^{(t)}_{\text{up}},\\
    -(\alpha^{(t)}_{i} - \rho^{(t)}_{\text{up}})                  & \text{if} \qquad \alpha^{(t)}_{i} > \rho^{(t)}_{\text{up}}.
\end{cases}
\label{eq:reward}
\]
Notably, the first case linearly penalizes the occurrence of under provisioning decisions, while the third case acts in a similar way on the over-provisioning cases. The middle case is the target scenario, which assumes correct PRB allocation decisions in response to the instantaneous slice traffic request.
We envision the multi-agent RAN slicing problem as a sequential procedure, where at the beginning of each decision interval $t$, the different agents perform local decisions according to a priority value $\mu_i$. Nevertheless, multiple and independent agents may perform inaccurate decisions and leave the subsequent agents with no spare resources, specially in the initial training phase. Therefore, at the end of each training period, we calculate a penalty
\begin{equation}
P_i^{(t)}=-\eta_i\mathds{1}\left(a_{i}^{(t)}  > \nu_{i}^{(t)} \right), 
\end{equation}
where $\eta_i$ is the penalty coefficient of the $i$-th slice, and $\mathds{1}$ denotes the logical operator. This penalty overrides the instantaneous agent reward $r_i$ if the decision $a_{i}^{(t)}$ is greater than the amount of spare resources left by the previous decisions of the other agents, that in turn prevents the agents to exceed the available resources at the base station.
This design choice is justified by the results provided in Sec.\ref{single_bs_scenario}.


\begin{figure}[t!]
\centering
\includegraphics[width=.7\columnwidth, trim={1cm 0 2cm 0},clip]{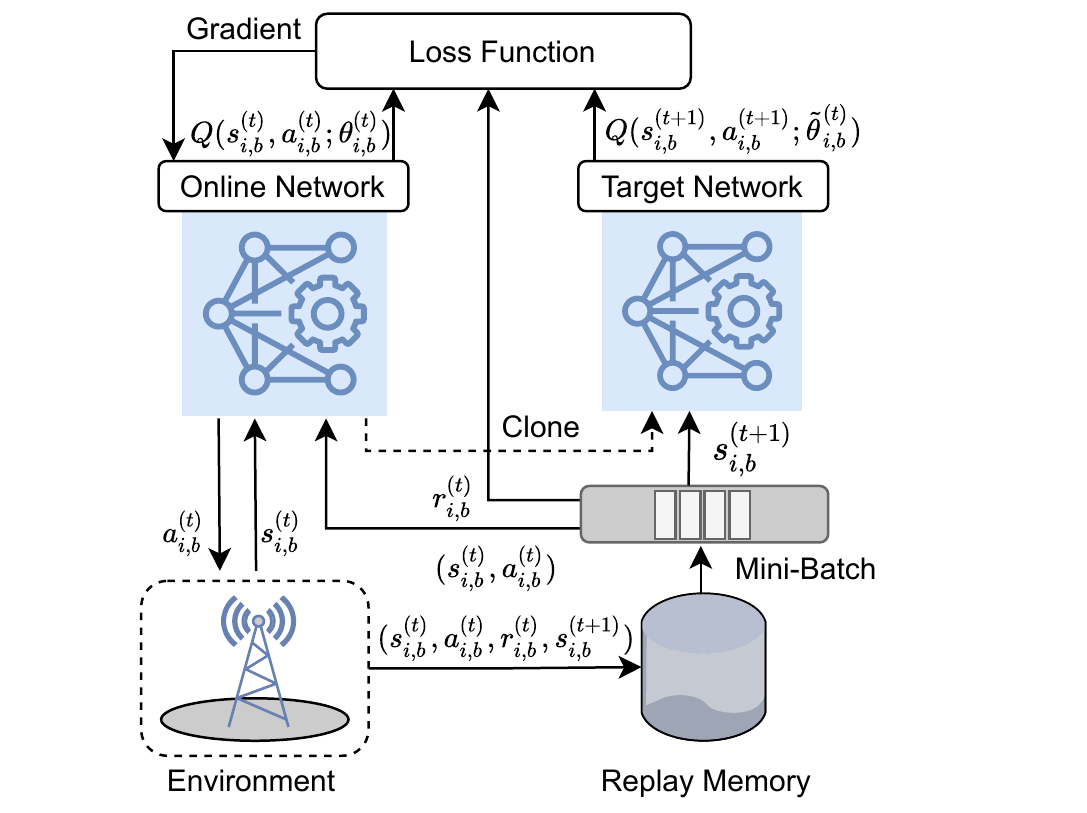}
\caption{ \small An illustration of DDQN workflow.}
\vspace{-6mm}
\label{fig:DDQN}
\end{figure}

\textbf{Training of Agents}
The training of the local agent implies the characterization of the action-value function $Q \colon \mathcal{S} \to \mathcal{A}$.
Let us define the policy $\pi$ as a probabilistic function mapping states to actions. The agent makes decisions and selects the corresponding actions based on $\pi$, determining the best action for each state. Under a given policy $\pi$, the action-value function can be defined as, $Q_{\pi}(s^{(t)}, a^{(t)}) = \mathbb{E}_{\pi}\left[\sum_{n=0}^{\infty}\left(\gamma^{n} r^{(t+n+1)} | (s^{(t)}, a^{(t)}) \right)\right]$, where $\gamma \in [0,1]$ is a discount factor that 
weights the short-sighted and far-sighted reward, and $n$ is the temporal index. According to Bellman's equation~\cite{Bellman}, the optimal state-action value function can be expressed as ${Q}^{\star}(s^{(t)},a^{(t)}) = \mathbb{E}\left[r^{(t)}+\gamma\underset{a^{(t+1)}}{\max}{Q}^{\star}{(s^{(t+1)},a^{(t+1)} | s^{(t)}, a^{(t)}}) \right]$,
and thereby the Q-learning update rule based on temporal difference (TD)~\cite{TD-error} is given by, 
\begin{multline}
Q(s^{(t)},a^{(t)}) \leftarrow
 Q(s^{(t)},a^{(t)}) +\\ \xi[r^{(t)}+\gamma \underset{a^{(t+1)}}{\max}Q(s^{(t+1)}, a^{(t+1)})
- Q(s^{(t)},a^{(t)})],
\end{multline}
where $\xi$ is the learning rate. DQN adopts deep neural network (DNN) to approximate the state-action value and surmount the curse of dimensionality concerning inordinate large state spaces. To limit the catastrophic interference problem~\cite{catastrophic_g}, which is the tendency of a neural network to forget about previously learned information upon learning new ones, we adopt an experience replay strategy. 
%
In particular, let us introduce $\beta_{i}$ as the experience buffer.
As depicted in Fig.~\ref{fig:DDQN}, in every training interval, we store the tuple ${({s}_{i}^{(t)},{a}_{i}^{(t)},{r}_{i}^{(t)},{s}_{i}^{(t+1)})}$ describing the instantaneous experience generated by the agent while interacting with the environment, and sample from $\beta_{i}$ a random batch of past experiences to regularize the training.

Additionally, DQNs are well known to provide an overoptimistic value estimation. We alleviate this problem by leveraging an additional DQN network, in the form of DDQN~\cite{DoubleDQN}-\cite{DDQN_VNF_OPT}. With a slight abuse of notation, let us introduce $Q(s_{i}^{(t)}, a_{i}^{(t)}; \theta_{i}^{(t)})$ and $Q(s_{i}^{(t)}, a_{i}^{(t)}; \tilde{\theta}_{i}^{(t)})$ as the online network and target network respectively, where $\theta_{i}^{(t)}$ and $\tilde{\theta}_{i}^{(t)}$ denote the model parameters. To optimize the parameter set $\theta_{i}^{(t)}$ and approximate the optimal action-value function ${Q}^{\star}(s_{i}^{(t)},a_{i}^{(t)})$, we use the following loss function,
\begin{equation}
    L(\theta_{i}^{(t)}) = \mathbb{E}[y_{i}^{(t)} - Q(s_{i}^{(t)}, a_{i}^{(t)}; \theta_{i}^{(t)})]^2,
\end{equation}
where $y_{i}^{(t)} = r_{i}^{(t)} + \gamma \underset{a^{(t+1)}_{i}}{\max}Q(s_{i}^{(t+1)}, a^{(t+1)}_{i}; \tilde{\theta}_{i}^{(t)})$ and $\tilde{\theta}_{i}^{(t)}$ is copied from $\theta_{i}^{(t)}$ at the end of each episode.
Finally, the objective function of the DDQN model can be written as,
\begin{equation}
    y_{i}^{(t)} = r_{i}^{(t)} + \gamma Q(s_{i}^{(t+1)}, \underset{a^{(t+1)}_{i}}{\argmax}Q(s_{i}^{(t+1)}, a^{(t+1)}_{i}; \theta_{i}^{(t)}); \tilde{\theta}_{i}^{(t)})),
\end{equation}
where $\theta_{i}^{(t)}$ is a local training model used for selecting actions, and $\tilde{\theta}_{i}^{(t)}$ is used to evaluate their values according to a different policy, thus mitigating over-estimations issues and improving the decision agents' performances~\cite{DoubleDQN}. The loss function estimates the difference between true action-value and target action-value. As the overall training procedure aims at minimizing this loss function, we adopt stochastic gradient descent (SGD) approach~\cite{Stochastic-gd} to pursue this goal.
The local agent training procedure is summarized in Algorithm~\ref{algo:DRL}. The overall local process is aided by a federation scheme (lines 1-6) described in details in the following subsection.

\begin{algorithm}[!t]
\small
\SetKwInOut{Input}{Input}
\SetKwInOut{Output}{Output}
\SetKwInOut{Return}{return}
\SetKwInOut{Initialize}{Initialize}
\Input{  $t, T, \hat{T}, i \in \mathcal{I}, \theta_{i,b}^{(t)}, \Omega_{k}^{(t)}$ \;}
\Output{Improved DDQN model $\theta_{i,b}^{(t+1)}$ \;}
\Initialize{ $\theta_{i,b}^{(0)}, \forall b \in \mathcal{B}, t = 0 $\; }
      \If{$mod(t,\hat{T})==0 \land t > 0$}{ 
        Upload $\theta_{i,b}^{(t)}$ \;
        Wait for Algorithm~\ref{algo:FDRL}\;
        \emph{ \#Get FL model and update the local one}\;  
        $\theta_{i,b}^{(t+1)} \leftarrow \Omega_{k}^{(t)}$\;
   } 
    \For{ $b \in \mathcal{B}$, in parallel}{
        $\nabla L(\theta_{i,b}^{(t)})  \leftarrow $ Local model training\;
        $\theta_{i,b}^{(t+1)} \leftarrow \theta_{i,b}^{(t)} + \nabla L(\theta_{i,b}^{(t)})$\; %
    } 
\caption{DRL RAN resource allocation for the $i$-th slice}
\label{algo:DRL}
\end{algorithm}
\setlength{\textfloatsep}{0.5pt}

\subsection{Federated DRL for RAN Slicing}
\label{sec:federated}
FL allows training machine learning models across multiple decentralized entities which have access to a limited set of the overall data available. Conversely to multi-agent reinforcement learning, which defines a set of autonomous agents that observe a global state (or partial state) of the system, select individual actions and receive individual rewards, FL allows to collaboratively learn a shared prediction model by iteratively aggregating multiple model updates, thus decoupling the learning procedure from the need of centralized data sources. A refined version of the original models, combination of multiple local models according to specific federation strategies, is then shared to the agents allowing to significantly improve the learning rate, ensure privacy~\cite{not_just_privacy} and provide better generalization~\cite{fl_Aled}.

As depicted in Fig.~\ref{fig:architecure}, within the context of our FDRL-based framework each agent trains a local DDQN model $\theta_{i,b}^{(t)}$ and shares its experience, under the form of model hyperparameters, to those entities belonging to the corresponding federation layer. This iterative training approach enables each federation layer to aggregate the collected knowledge of single agents into a global updated model $\Omega_{i}^{(t+1)}$, usually stored into a cloud platform or a nearby edge platform to allow faster feedbacks.
In order to enhance efficiency and avoid communication overhead, we allow the federation layer to collect the local models (and share the updated ones) only every $\hat{T}$ decision intervals, defining this time period as \emph{federation episode}.
Different strategies can be adopted to derive the global federated model, each one implementing a predefined federation strategy function $f_{\textit{strategy}}(\cdot)$.

In \emph{Average} federation strategy, dubbed as \emph{FDRL} in the following of this work, the collective federation model for the next time interval $\Omega_{i}^{(t+1)}$ is derived as the simple average of the incoming model weights belonging to all the agents, as
\begin{equation}
    \Omega_{i}^{(t+1)} = \frac{1}{B}\sum\limits_{b \in \mathcal{B}}\theta_{i,b}^{(t)}.
    \label{eq:averaging}
\end{equation}
Aggregated mobile traffic demands follow repetitive spatio-temporal trends due to human activities~\cite{Spatio-LTM}. 
In this context, it is expected that a good characterization of such processes would allow more accurate forecasting of the network utilization and, in turn, enable an efficient and even proactive planning of the resource allocation.
However, as highlighted in~\cite{DeepCogInfocom19}, it is not enough to leverage the geographical locations and related spatial proximity of the BS to obtain a comprehensive view of traffic demands, as the land usage of the slice resources may differ even within base stations belonging to the same geographical areas. This introduces an additional issue in our framework, as not all the federated agents should exchange knowledge with each other, nor this should be restricted to only nearby entities. To address this fundamental issue, in the following we propose a clustering algorithm to guide DA subsets definition, based on network monitoring traces and their similarity.

\begin{figure}[h]
\centering
\includegraphics[clip, trim = 0cm 0cm 0cm 2cm, width=\columnwidth]{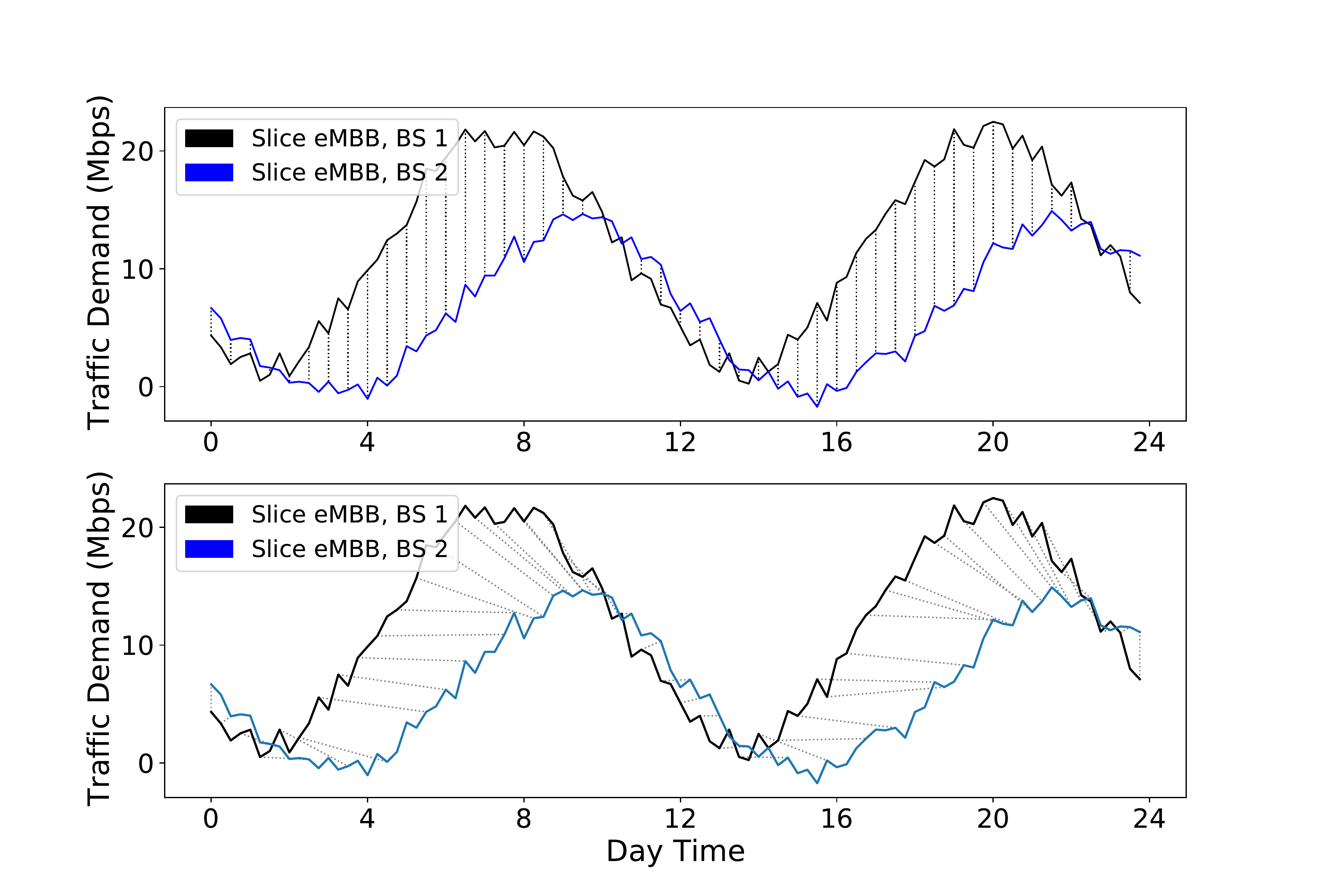}
\caption{\small Example comparison of Euclidean distance against Dynamic Time Warping distance over traffic demand time series.}
\label{fig:DTW}
\end{figure}

\subsection{Dynamic Traffic-Aware Agent Selection}
\label{subsec:dynamic_clustering}
Given the rapid spatio-temporal variation of the traffic demand due to end-user mobility, we advocate for the setup of a clustering algorithm to derive the subset of slice agents that should exchange their local knowledge, while considering both mobility and traffic demand variations.
Let us introduce $\tau_{i,b}$ as the time series describing the downlink traffic demand of the $i$-th slice instantiated over base station $b$.
Then, for each pair $j,z\in \mathcal{B}$, we can compute the similarity of the recorded monitoring information as $DTW^{(j,z)}=f_{DTW}(\tau_{i,j},\tau_{i,z})$, where $f_{DTW} (\cdot)$ is the Dynamic Time Warping distance~\cite{DTW}, a state-of-the-art distance metric for time series analysis\footnote{We refer the reader to~\cite{DTW} for an exhaustive explanation.}.
DTW is particularly suitable in our scenario as it allows, conversely to standard distance metrics, e.g., Euclidean distance, to calculate accurate similarity value even in presence of differently sized sequences, and independently of their time shift. An example of DTW distance calculation is depicted in Fig.~\ref{fig:DTW}, where it can be noticed how maximum and minimum values of the traces are correctly mapped to each other. The pairwise distances are then collected into the distance matrix $\mathbf{D}= (DTW^{(j,z)}) \in \mathbb{R}^{|\mathcal{B}| \times |\mathcal{B}|}$, and provided as input of our clustering algorithm.
DTW has linear space complexity, but quadratic time complexity. To reduce the latter, a number of options are available. In our case, we limit the maximal shift by setting a fixed time a window of few hours, thus reducing the complexity even in case of long sequences. Nevertheless, recent work from~\cite{DTW_Barrier} proposed a novel efficient implementation which breaks the quadratic time complexity to $O(n^2 \log{n})$, where $n$ is the length of the sequence. 
To perform the final classification, we rely on an extended version of the Density-based spatial clustering of applications with noise (DBSCAN) algorithm, introduced in~\cite{DBSCAN}. DBSCAN is a non-parametric density-based clustering algorithm that allows finding the most representative points within a dataset (also known as \emph{core samples}) based on their density in a multi-dimensional space, and expands clusters from them. It expects two inputs: $\epsilon_d$, representing the maximum distance between two samples for one to be considered as in the neighborhood of the other, and $n_{min}$, which defines the minimum number of samples in a neighborhood of a point to be considered as a core sample.

\begin{algorithm}[!t]
\small
\SetKwInOut{Input}{Input}
\SetKwInOut{Output}{Output}
\SetKwInOut{Return}{return}
\Input{$t, T, \theta_{i,b}^{(t)}$ $\forall b \in \mathcal{B}$, $\tau_{i,b}^{(t)}$ $\forall b \in \mathcal{B}$, $\epsilon_d$, $n_{min}$}
\Output{Improved federation models $\Omega_{i,k}^{(t+1)}, \forall \Psi_k \in \Psi$}
\emph{\#Define clusters and send initial/updated model}\;
        Collect $\tau_{i,b}^{(t)}$, $\forall b \in \mathcal{B}$ \;
        Compute $\mathbf{D} = (DTW^{(j,z)}), \forall j,z \in \mathcal{B}$\; 
        $\Psi_k \in \Psi \leftarrow DBSCAN( \mathbf{D},\epsilon_d,n_{min})$\;
 \While{ $t < T$}{
  \If{$mod(t,\hat{T})==0 \land t > 0$}{
          \For{ each $\Psi_k \in \Psi$, in parallel}{
                Collect $\theta_{i,k}^{(t)}, \forall b \in \Psi_k$\;
                \emph{ \#Derive FL models based on Fed. strategy}\;  
                $\Omega_{i,k}^{(t+1)} \leftarrow f_{\textit{strategy}}(\theta_{i,b}^{(t)}, \forall b \in \Psi_k)$\;
                $\theta_{i,b}^{(t+1)} \leftarrow \Omega_{i,k}^{(t+1)}$ $\forall b \in \Psi_k$\;
         }
        \emph{ \#Return updated local models}\;  \
        \Return{$\theta_{i,b}^{(t+1)}, \forall b \in \mathcal{B}$}
    }
    Run Algorithm~\ref{algo:DRL}\;
 }
\caption{RAN resource orchestration for the $i$-th federation layer}
\label{algo:FDRL}
\end{algorithm}
\setlength{\textfloatsep}{0.5pt}

Given the above, at the end of each federation episode, we can derive in a dynamic way (and based on updated mobile monitoring information) the clusters $\Psi_k \in \Psi$, $k=\{1,\dots,|\mathcal{I}|\}$, where $\Psi$ is the cluster set. Each cluster includes the set of base station $b \in \Psi_k$ that should be involved in the following model update procedure. Therefore, the framework spawns multiple federation models $\Omega_k$, one for each detected cluster $k$, which evolve in parallel till the next federation episode. The pseudocode of our FDRL-based approach for RAN slicing resource orchestration is listed in Algorithm~\ref{algo:FDRL}.
We remark that in our framework multiple instances of Algorithm ~\ref{algo:FDRL}, i.e., one for each slice $i\in\mathcal{I}$, are deployed to build the corresponding FL domain for a given federation strategy $f_\textit{strategy}(\cdot)$. It follows that the updated federation model, combination of the information coming from the elements of the cluster $\Psi_k$ (described in line 10), can be derived following the \texttt{Full-Cluster} strategy, namely $f_{\textit{FC}}(\cdot)$, as:
\begin{equation}
    \Omega_{i,k}^{(t+1)} = \frac{1}{|\psi_k|}\sum\limits_{b \in \psi_k}\omega_{i,b}^{(t)}\theta_{i,b}^{(t)}, \qquad \forall \psi_k \in \psi
\end{equation}
where $|\psi_k|$ is the cardinality of $\psi_k$ , and $\omega_{i,b}^{(t)} = \frac{\hat{r}_{i,b}^{(t)}}{\sum\limits_{b \in \psi_k} \hat{r}_{i,b}^{(t)}}$ is a weight parameter.
It should be noted that within these settings, the federation step will occur among models with high degree of similarity, thus favoring the \emph{specialization} of the agents towards the most common traffic statistics.

Other complementary approaches can be defined to guide the agent selection and subsequent federation model update. In particular, upon the definition of the cluster set $\Psi$, we introduce \texttt{Random Representative} strategy $f_{\textit{RR}}(\cdot)$ as a baseline approach, which randomly selects a representative from each cluster:
\begin{equation}
    \psi_{\textit{random}} = \{x | x = \textit{rand}(\psi_k) , \qquad \forall \psi_k \in \psi\}
\end{equation}
and consequently defines the updated federated model as:
\begin{equation}
    \Omega_{i}^{(t+1)} = \frac{1}{|\psi_{\textit{random}}|}\sum\limits_{b \in \psi_{\textit{random}}}\omega_{i,b}^{(t)}\theta_{i,b}^{(t)}.
\end{equation}
Similarly, let us introduce the \texttt{Best Representative} strategy, as a method that derives the updated federation model by selecting a representative agent from each cluster as follows:
\begin{equation}
    \psi_{\textit{best}} = \{x | x = \underset{k}{\argmax R_k} , \qquad \forall \psi_k \in \psi\}
\end{equation}
where $R_k$ is the cumulative reward within the past federation episode.
Thus, the model update strategy \texttt{Best Representative} $f_{\textit{BR}}(\cdot)$, can be defined as:
\begin{equation}
    \Omega_{i}^{(t+1)} = \frac{1}{|\psi_{\textit{best}}|}\sum\limits_{b \in \psi_{\textit{best}}}\omega_{i,b}^{(t)}\theta_{i,b}^{(t)}.
\end{equation}
By combining single models derived from each cluster, we can pursue higher \emph{generalization} of performances, i.e., aim at a federated model able to deal with heterogeneous traffic statistics.

\begin{figure}[t]
\centering
\includegraphics[width=0.7\columnwidth]{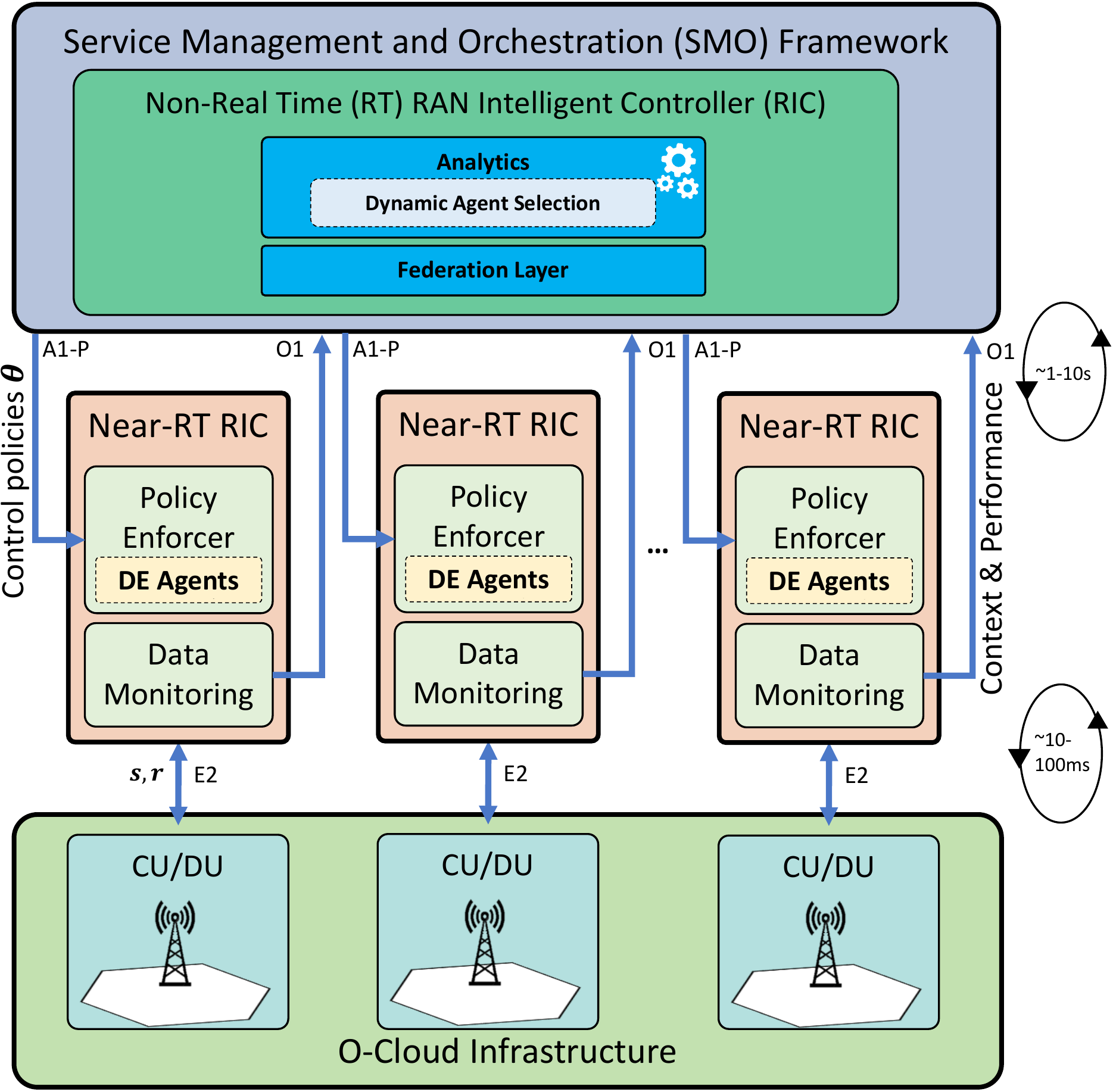}
\caption{\small O-RAN compliant system architecture.}
\label{fig:ORAN}
\end{figure}
\section{O-RAN Compliance}
\label{sec:ORAN}

The design of our solution closely follows the O-RAN framework~\cite{ORAN_architecture}.
O-RAN represents a worldwide effort to reach new levels of openness in next-generation virtualized radio access networks (vRANs). Driven by major carriers, it aims at disrupting the vRAN ecosystem traditionally dominated by a small set of player by breaking vendors' lock-in and opening the business market~\cite{ORAN_NEC}.
The most important functional components introduced by O-RAN are the non-real-time (non-RT) radio intelligent controller (RIC) and the near-RT RIC~\cite{ORAN-infocom}.
The main functionality provided by the Non-RT RIC it to support RAN optimization over relatively large time scales (e.g., seconds or minutes). This often implies machine learning (ML) model training and subsequent control policy definition, to be enforced via the A1 interface towards the distributed Near-RT RICs.
The Near-RT RIC is a logical function that enables near-real-time optimization and control, as well as data monitoring of O-RAN central unit (O-CU) and O-RAN distributed unit (O-DU) nodes (which support eNBs/gNBs deployment as virtualized network functions (VNFs)) in near-RT timescales (between 10 ms and 1 s).
Fig.~\ref{fig:ORAN} depicts a high-level view of the O-RAN architecture, highlighting the synergies with respect to our proposed approach. In particular, we envision our federated learning and dynamic agent selection module as co-located with Non-RT RIC, which handles the A1's Policy Management Service to enforce radio policies.
On the other side, local agents co-located with the Near-RT RICs collect this information, perform local decisions, and exploit the E2 interface to forward resulting radio policies to the base station. The same E2 interface would allow the local agent to gather base station KPIs for the purpose of model training and monitoring.

\begin{figure}[t]
\centering
\includegraphics[clip, trim = 0cm 0cm 0cm 0cm,width=6cm]{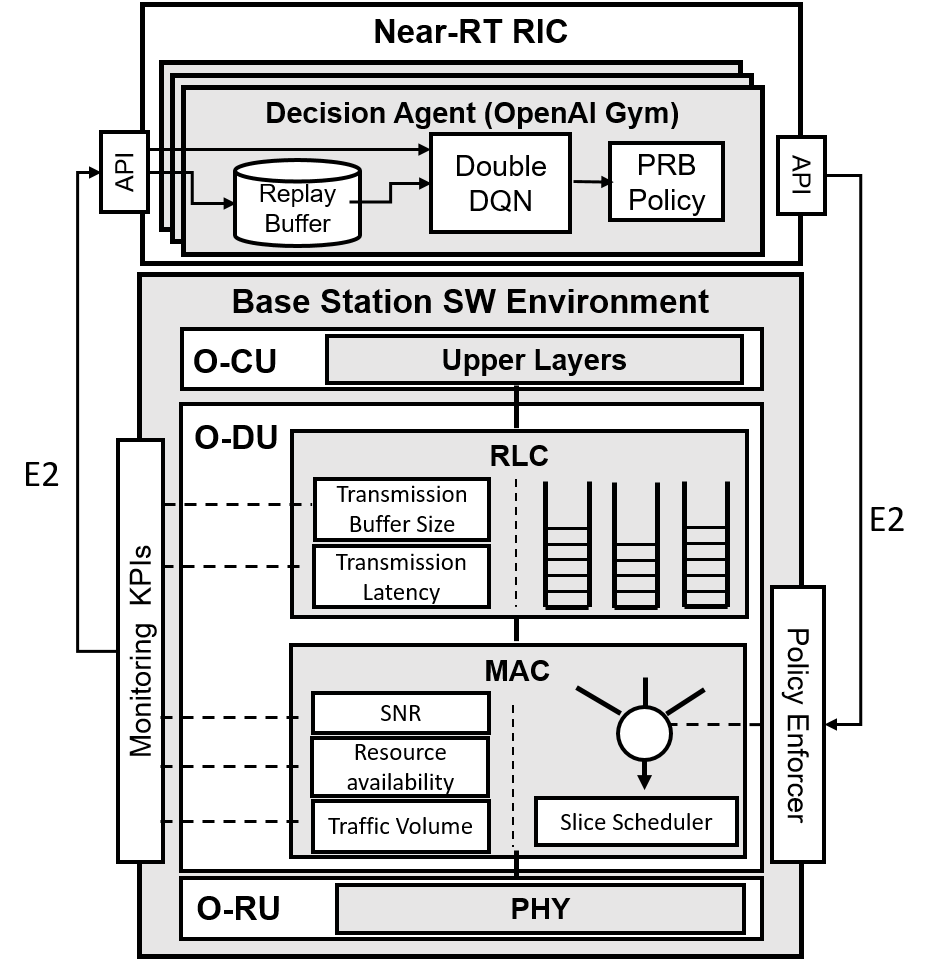}
\caption{\small Software architecture and protocol stack overview.}
\label{fig:sw_architecture}
\end{figure}

\change{As depicted in Fig.~\ref{fig:sw_architecture}, we implement our framework in Python programming language, exploiting OpenAI Gym library~\cite{Gym_Broc} and interfacing DRL agents with a custom base station simulator environment, which includes virtual transmission queues and main PHY/MAC/RLC functionalities, together with O-RAN E2 interface to allow gathering the slice networking statistics from each distributed unit (O-DU), and to enforce PRB policy decisions in the BS slice scheduler based on defined state space and action space in Sec.~\ref{sec:DQN}. Finally, as described in Sec.~\ref{sec:federated}, a federation layer connects the DRL agents of the $i$-th slice to enable inter-agent information exchange and expedite the overall learning procedure. The procedure is summarized in Algorithm~\ref{algo:DRL} and~\ref{algo:FDRL}. }

\section{Performance Evaluation}
\label{sec:perf_eval}
In this section, we evaluate our proposed architecture  numerical simulations on a dedicated server, equipped with two Intel(R) Xeon(R) Gold 5218 CPUs @ 2.30GHz and two NVIDIA GeForce RTX 2080 Ti GPUs. Moreover, the DNNs are implemented based on TensorFlow-gpu version 2.5.0. In neural network architecture, we use two fully connected layers with $24$ neurons activated by ReLU function for each layer where the target network is updated per episode and each episode consists of $5$ decision intervals, or epochs. Each decision interval has a duration of $\epsilon = 60$ seconds, during which local monitoring information is collected to build the local agent state. Online and Target networks are characterized by the same DNN structure. 
The hyperparameter tuning depends highly on capability, scenario, and technology used~\cite{Globe_far}. The network parameters are updated using the Adam optimizer~\cite{ADAM}. The discount factor $\gamma$ and learning rate $\xi$ are set to be $0.99$ and $0.001$ respectively. The replay buffer size of each agent $\beta_{i,b}$ is set to $20000$ samples, out of which a batch of $32$ samples is extracted for each training interval. Without loss of generality, We set $\eta_i=100$ as penalty value for all the slices. In order to provide a comprehensive overview, we first evaluate single base station settings, focusing on the capabilities of single agents to deal with RAN resource allocation. Then, we address a more realistic scenario considering a multi-slice deployment over several RAN nodes, accounting for end-user mobility and variable traffic demands.

\subsection{Local Agent Performance Assessment}
\label{single_bs_scenario}
In our proposed framework, DRL agents optimally allocate radio resources to each slice, while a federation layer enables a periodical exchange of the DRL's parameter values to improve the learning process across multiple agents of the same slice.
\begin{figure}[t]
\centering
\includegraphics[clip, trim = 0cm 0cm 0cm 0cm,width=\columnwidth]{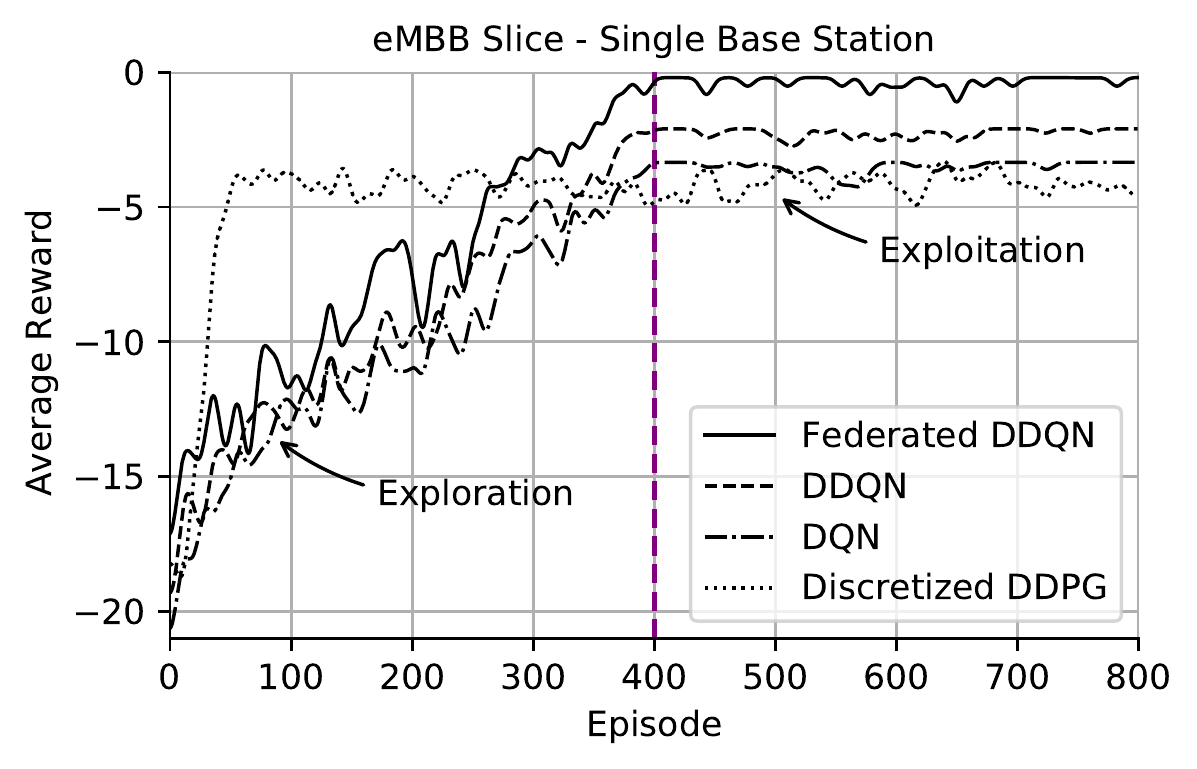}
\caption{\small The convergence performance of different local decision algorithms and an FDRL approach for a single decision agent.} 
\label{fig:Local_Model_Perf}
\end{figure}
\begin{figure*}[!t]
\centering
\includegraphics[clip, trim = 0cm 0cm 0cm 0cm, width=0.85\textwidth]{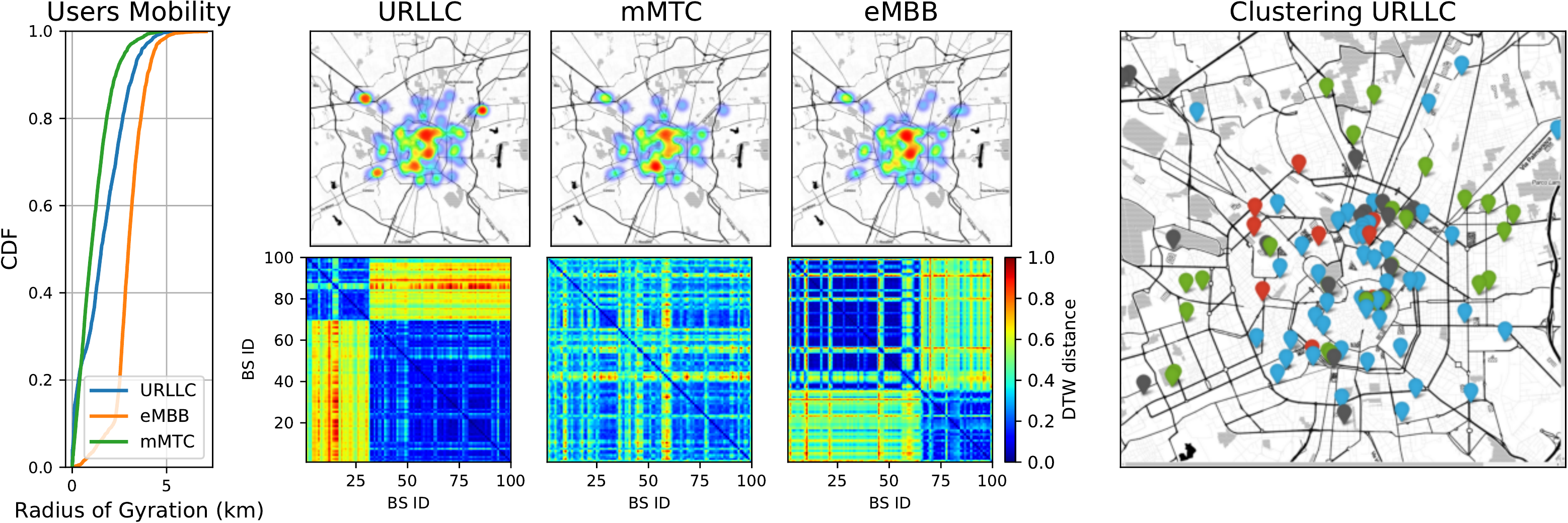}
\caption{\small Mobility Statistics for different network slices. (Left) CDF of end-users radius of gyration, (Center) Average Spatial distribution of slice users over 24h time span, and corresponding DTW distance matrix. (Right) Example of resulting clustering for the URLLC slice case, each color defines a different cluster.}
\label{fig:Mobility}
\end{figure*}
First, we compare the performances of different RL algorithms when dealing with radio resource allocation, without involving federated learning. To this aim, we consider a base station scenario including $3$ network slices, i.e., an ultra reliable low latency communication (URLLC) kind of slice, an enhanced mobile broadband (eMBB), and one dedicated to massive machine-type communications (mMTC) traffic, each one characterized by the SLA latency values of $\Lambda_i = [10, 40, 20]$ ms, respectively~\cite{GAN_Powered}.
Regarding the throughput requirements, we do not assume any fixed value as it would depend on the random mobility pattern of the end users and their generated traffic. Instead, we enforce (with a slight abuse of notation), $\lambda_{i,b}^{(t)} = \varphi_{i,b}^{(t)}$ for every decision interval to let the agents adapt their decisions to the instantaneous traffic volumes.
We instantiate a DA in every base station for each slice. We model the instantaneous traffic demand of each slice as the realization of a Poisson distribution with mean value $\lambda_i$, and emulate the SNR variability extracting its instantaneous values from a Rayleigh distribution with the average value set to $25$ dB. Moreover, we set $\iota = 10$ PRBs as the minimum resource allocation step.
Fig.~\ref{fig:Local_Model_Perf} depicts the training procedure for the eMBB slice, comparing different local decision algorithms. In particular, we consider the single DQN approach, which implements standard Q-Learning procedures, discretized Deep Deterministic Policy Gradient (d-DDPG) a popular reinforcement learning algorithm~\cite{DDPG}, and our DDQN scheme. 

We let the scenario run for $800$ federation episodes, and depict the results in terms of cumulative reward, as defined in \change{Eq.~\eqref{eq:reward}}.
The variability of the network slicing environment leads to experience learning curves with high fluctuations. For visual clarity, results are averaged over $10$ simulations. As expected, the DQN approach hardly copes with the definition of suitable PRB allocation policies, providing lower performances both in terms of cumulative reward and convergence time. Similarly, d-DDPG suffers the temporal periodicity of the traffic demand, resulting in a steep learning curve that soon saturates to suboptimal performances. 
Conversely, after an initial exploration phase, the DDQN approach is able to allocate in a more consistent way correct amount of PRBs to each slice according to the corresponding real-time traffic and latency demands. It is worth highlighting that in terms of convergence time, in general, FDRL schemes do not necessarily provide better performances when compared against standard DRL approaches. In fact, one of the main features of FL is that it allows local DRL agents to indirectly gain knowledge on a wider state space, extending the local experience with that coming from other decision entities deployed within the same environment. This enables the DAs to provide more robust performances when deployed in realistic environments. Nevertheless, the same Fig.~\ref{fig:Local_Model_Perf} provides an overview of the local model training procedure, with and without the adoption of FL schemes. In our considered scenario, it can be noticed how DRL curves (dashed lines in the plot) present slower convergence time and higher fluctuations when compared against Federated DDQN approach (solid line in the plot).
Additionally, DRL curves present lower cumulative reward after $400$ episodes, suggesting a lower capability of the DAs to adapt their decisions at the fast-changing network slicing environment considered in our work.

\subsection{System-level Simulations}
\label{multi_bs_scenario}

\subsubsection{Mobility and Traffic Demand Characterization}
\begin{figure}[t]
\centering
\includegraphics[clip, trim = 0cm 0cm 0cm 0cm, width=.95\linewidth]{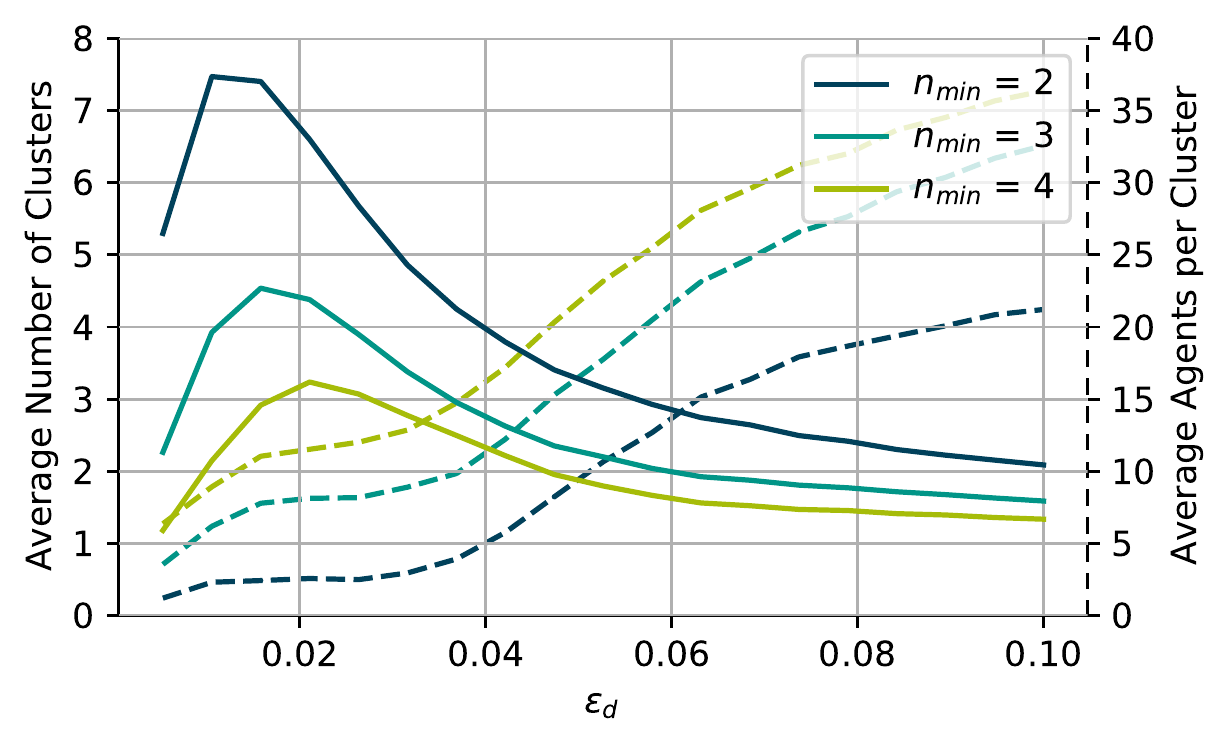}
\caption{\small Sensitivity analysis performed on the clustering parameters and generated traffic traces.}
\label{fig:clustering_hyperparameter}
\end{figure}

In order to validate our framework in realistic settings, we consider the city of Milan, Italy, as scenario of study. We collect city-wide RAN deployment information including more than $50$ BSs from publicly available sources\footnote{https://opencellid.org/}, and simulate realistic human mobility patterns leveraging the work of~\cite{Mobility_model}.
The density exploration and preferential return (d-EPR) algorithm allows capturing mobility patterns by specifying as input the geographical position of the base stations together with several probabilistic parameters. We let the model evolve adopting the default parameters described in~\cite{dEPR}. By defining the location relevance on the mobility space, we can influence the next-hop selection of each end-user, therefore emulating a higher concentration of mobile devices in specific areas of the city over time, e.g., the daily commuting over the city center during working days.
Fig.~\ref{fig:Mobility} (Left) depicts the CDF of the resulting radius of gyration per slice, aggregating the results over $15000$ end-users equally distributed among the different slices. 
Without loss of generality, we consider the set of BSs characterized by the same radio capacity $C_b=100$ PRBs, and assume the same $3$ slices introduced above simultaneously running over all the BSs. 
Fig.~\ref{fig:Mobility} (Center) depicts the resulting spatial distribution of the end-users, accounting for a temporal time span of a full day.
From the picture, it can be noticed how the spatial distribution of slice users is actually similar along with the slice set, and influenced in specific areas of the city by the high density of RAN nodes. This is due to the d-EPR algorithm, which favors the next-hop destination of each user to happen towards a nearby point of interest, or, in our settings, the closest base station location.
The instantaneous traffic demand of each end-user is derived starting from the values reported in Sec.\ref{single_bs_scenario}
, weighted by a temporal factor to account for the traffic demand fluctuations typical of mobile network scenarios, as those presented for example by~\cite{pi_ROAD} and~\cite{Wang_18}.
In the lower part of Fig.~\ref{fig:Mobility}, we depict the resulting distance matrix $\mathbf{D}$ of each, i.e., per slice, downlink traffic demand, calculated at the beginning of every federation episode for each base station pair over the past $24$ hours. As detailed in Sec.~\ref{subsec:dynamic_clustering}, this matrix is used as input to an instance of the DBSCAN algorithm to derive the set of DAs (belonging to the $i$-th slice) which should be involved in the next federation episode and model exchange. Fig.~\ref{fig:Mobility} (Right) shows the resulting output clustering for the URLLC slice case, using $\epsilon_d =0.06$ and $n_{min}=2$ as parameters. Such values have been empirically selected following the sensitivity analysis depicted in Fig.~\ref{fig:clustering_hyperparameter}, which certifies that along the evaluation timeline and across the different running slices, the selection algorithm identified on average $3$ clusters populated by $15$ agents each. The resulting behavior of DAs is heavily affected by the entities participating in the federation process. Therefore, such kind of characterization is fundamental to ensure performances.


\begin{figure}[h]
\centering
\includegraphics[clip, trim = 0cm 0.2cm 0cm 0cm,width=\columnwidth]{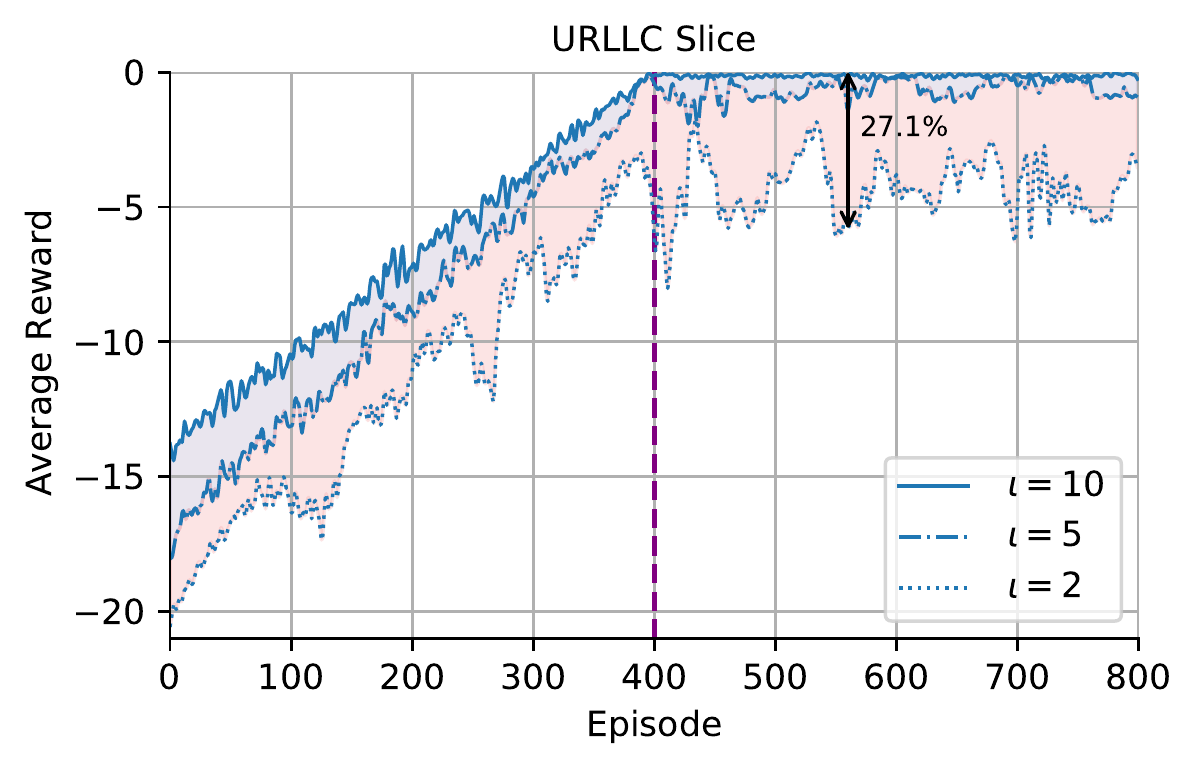}
\caption{\small Learning curve for different action spaces.}
\label{fig:PRB_Action_Space}
\end{figure}

\subsubsection{Effects of Different PRB Action space}

The size of the action space is well-known to affect the learning curve of any reinforcement learning algorithm. In Fig.~\ref{fig:PRB_Action_Space}, we investigate this aspect by varying the minimum PRB chunk size $\iota = {2, 5, 10}$ of the URLLC slice, while fixing $\hat{T}=5$ decision epochs per federation episode and adopting the full-cluster federation strategy. The plot shows how increasing the PRB chunk size, i.e., adopting smaller action spaces, actually influences the reward of the URLLC slice type and its stringent SLA requirements, with larger PRB chunk values achieving satisfactory performances in a faster way, with about 25\% performance gap with respect to $\iota = 2$. Nevertheless, a too broad PRB chunk allocation may result in resource wastage, with portions of the radio resources being under-utilized by the running slices. Such a trade-off should be carefully investigated according to both slice and system requirements. In the following, we will adopt $\iota=10$ PRBs whenever not specified otherwise.

\begin{figure}[h]
\centering
\includegraphics[clip, trim = 0cm 0.2cm 0cm 0cm,width=\columnwidth]{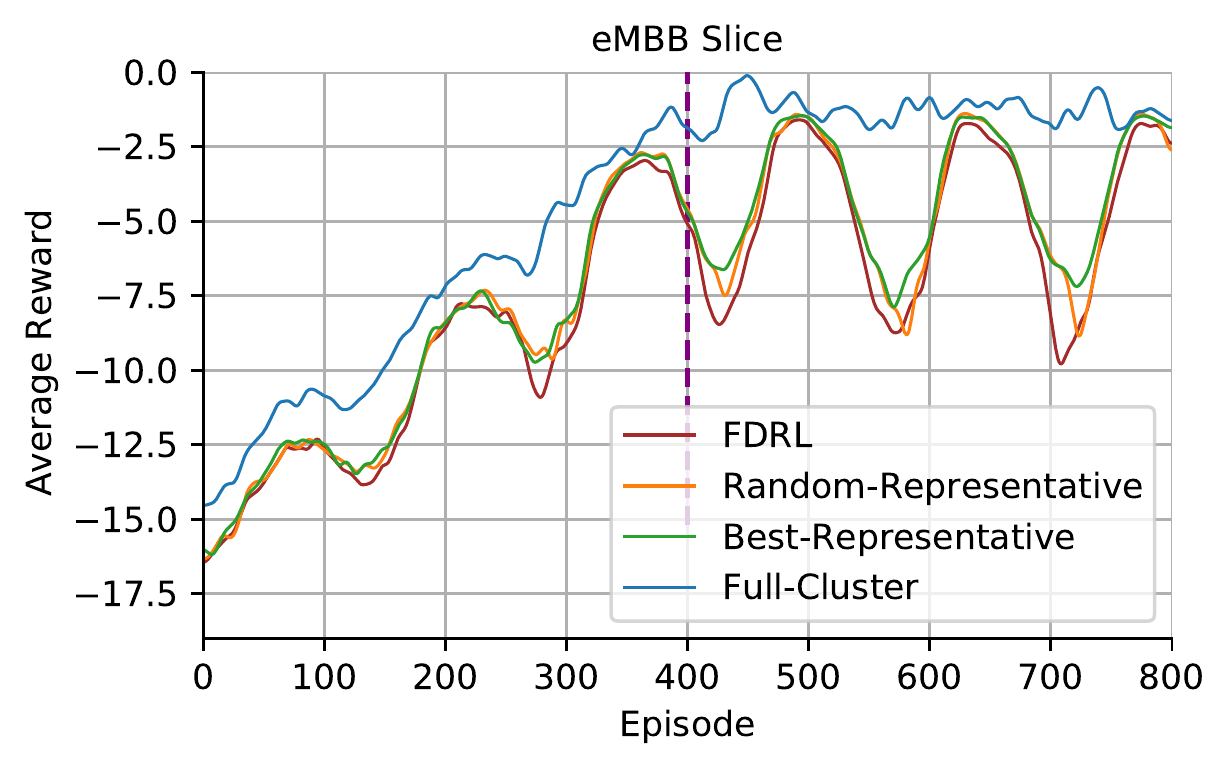}
\caption{\small Comparison of global performances for different dynamic and non-dynamic federation approaches. } 
\label{fig:Comparison}
\end{figure}

\subsubsection{Comparison of Different Federation Strategies}
Given the particular nature of the network slicing scenario, in this paper we advocate for a dynamic agent selection method based on the time similarity of traffic demands, dubbed as \emph{Dynamic Clustering} (DC). As discussed in Sec.~\ref{subsec:dynamic_clustering}, several strategies can be adopted to combine local models into federated ones, pursuing generalization and performance improvements.
In this paper we consider three DC aided approaches, namely \emph{Full-Cluster} (FC), \emph{Best-Representative} (BR) and \emph{Random-Representative} (RR), and compare their performances against a standard strategy which simply derives a new federated model accounting for all the available local models, without adopting any dynamic agent selection scheme, dubbed as \emph{FDRL}. The benchmark \emph{FDRL} approach exploits all the local trained models and the respective knowledge from the agents, and would theoretically allow for the best generalization of performances~\cite{FDRL_Generalization}. Interestingly instead, from our experiments it turns out that aggregation of widely heterogeneous local models actually limits the capability of the global federated model to converge to a one-fits-all unified model, motivating our dynamic agent selection approach which favors the specialization of federated agents working under similar RAN and mobility contexts.
Fig.~\ref{fig:Comparison} provides a comparison of learning performances for different federation strategies in terms of average reward and for $\hat{T}=5$. The agent's action selection follows a greedy approach which balances exploration of new actions and exploitation of already known decision policies.
We gradually limit the exploration capabilities in favour of the adoption of the learned policies, such that around half of the overall simulated time span, the possibility that the agent will explore new actions given a known instantaneous context is in the order of 2\%.
From the figure, we can observe how \emph{Full-Cluster} approach achieves better generalization of the learning policies, resulting in stable performances. Conversely, \emph{Best-Representative}, \emph{Random-Representative} and \emph{FDRL} suffer the dynamic behavior of the underlying traffic conditions, presenting inconsistent reward traces.

\begin{figure}[h!]
\centering
\includegraphics[clip, trim = 0cm 0.2cm 0cm 0cm,width=\columnwidth]{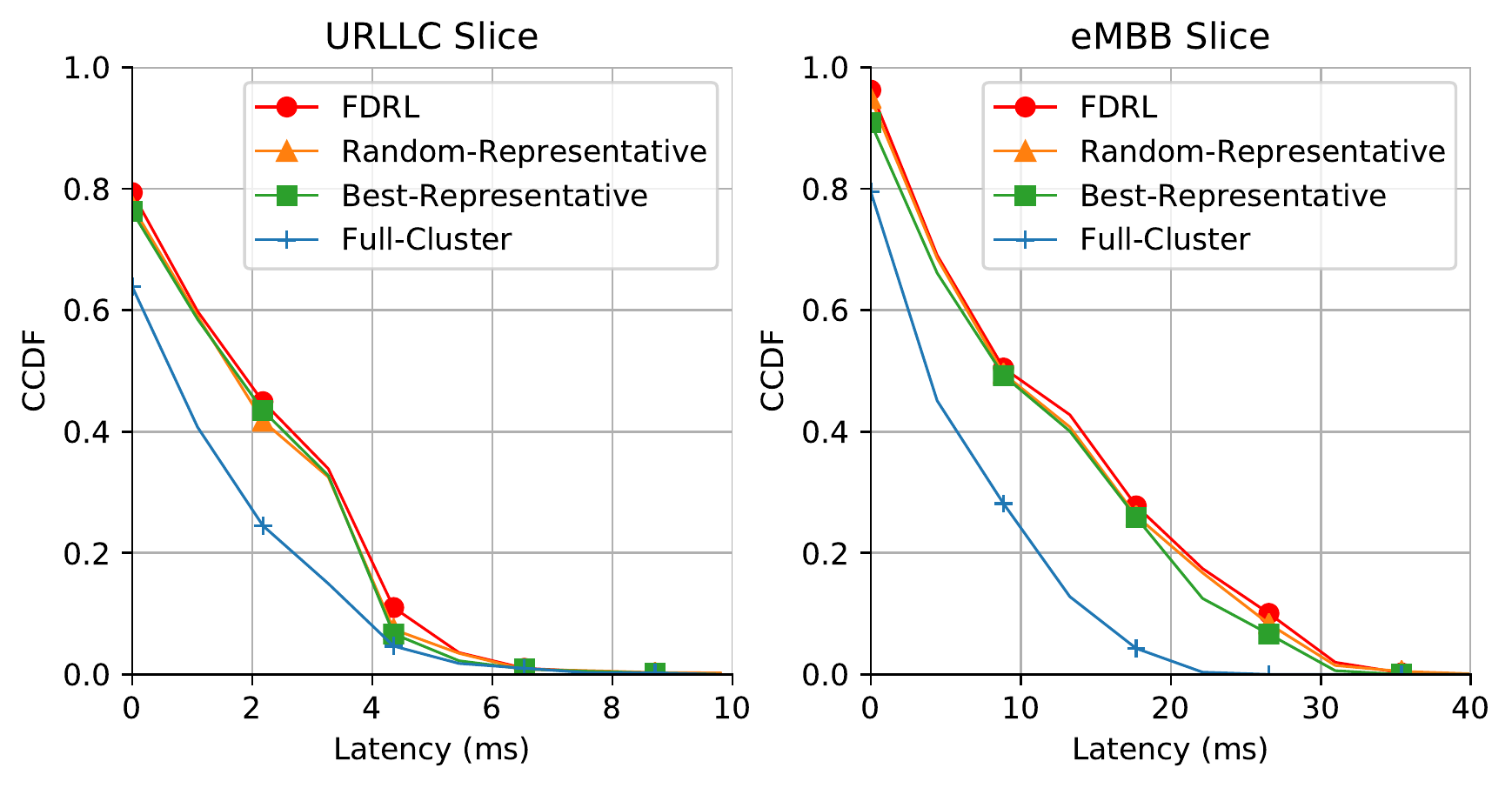}
\caption{\small \change{CCDF of transmission latency for URLLC and eMBB slices.}} 
\label{fig:Delay_sta}
\end{figure}
\subsubsection{\change{Latency Analysis for Different Federation Strategies}}
\change{

We continue our performance evaluation by considering the experienced transmission latency. We recall that as mentioned in Sec.~\ref{sec:scenario}, we define latency as the time spent by the slice traffic within transmission buffer of the base station. Fig.~\ref{fig:Delay_sta} depicts the complementary cumulative distribution function (CCDF) \cite{ccdf_ref}~\cite{ccdf_ref2} of the latency experienced by the URLLC and eMBB slices, resulting by different federation strategies.
\begin{figure*}[t!]
\centering
\includegraphics[clip, trim = 0cm 0.2cm 0cm 0cm,width=2\columnwidth]{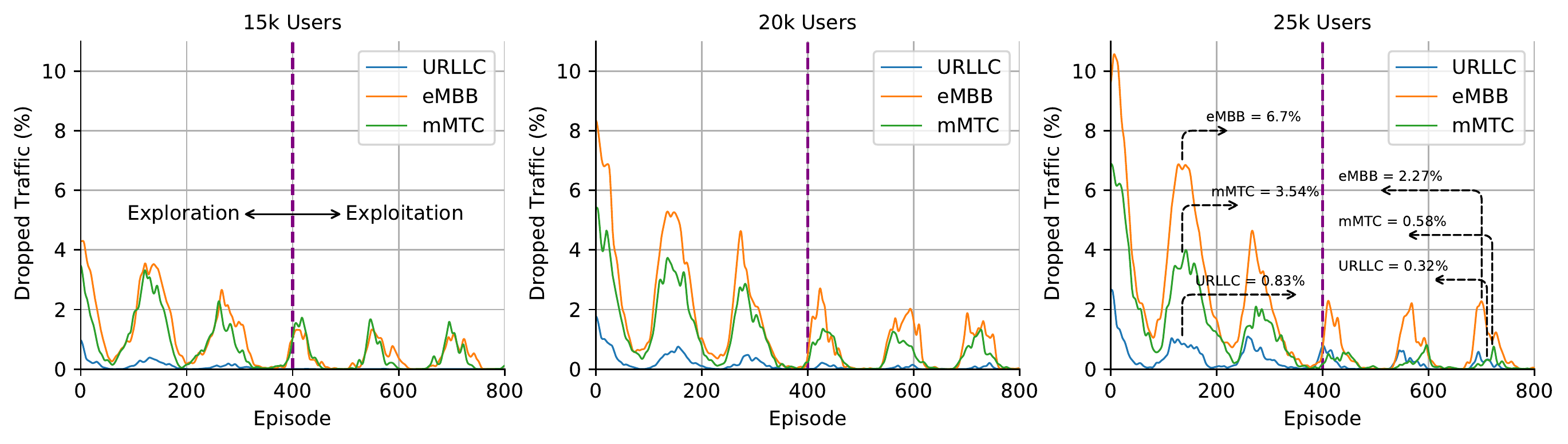}
\caption{\small Performance evaluation for different network loads derived by an increasing number of end-users in Full-Cluster settings.} 
\label{fig:network_load}
\end{figure*}
For the both slices, this latency is directly proportional to the traffic demand and the degree of contention of resources among the different slices, as well as to the resource allocation decisions taken by the agents. From the results, it can be noticed that the FDRL strategy leads to the worst performances, as having all the BSs involved in the learning process results in a slow adaptation of the decisions of the agents to the local traffic conditions, therefore leading to sub-optimal resource allocation and higher latency. 
In contrast, Full-Cluster presents a good trade-off in terms of collaboration among agents and specialization to the local traffic conditions, resulting in a more efficient PRB allocation and lower perceived latency. 
Finally, RR and BR federation strategies achieve performances comparable with the FDRL method, resulting from the limited cooperation in learning that leads these federation approaches to suffer more from the dynamic behavior of the underlying traffic conditions.}

\subsubsection{Effects of Different Network Loads and Mobility}
We continue our analysis investigating the performances of the \emph{Full-Cluster} method in heterogeneous traffic conditions. To this aim, we generate traffic and mobility dataset for an increasing number of end-users, namely 15k, 20k, and 25k.
As highlighted in~\cite{ARENA}, a non-linear relationship characterizes end-user mobility and throughput performances in crowded scenarios. Clearly, this also affects the communication latency, as a higher number of users will be simultaneously active under the same radio access node. In the context of RAN slicing resource allocation, this translates to finding the best DA logic to efficiently address such variability. In Fig.~\ref{fig:network_load}, we focus our analysis on the dropped traffic, i.e., the volume of traffic that did not meet the latency requirements due to wrong PRB allocation decisions, measured in percentage of the offered traffic volume of each federation episode.
From the picture, we can notice how during the initial exploration phase inexperienced PRB allocation decisions performed by the DAs heavily affect the latency requirements of all network slices, with peaks of dropped traffic that increase with the growing number of end-users. Nevertheless, this trend improves over time as the agents gain knowledge over the underlying scenario and get trained, finally converging after policy switch, i.e. after episode 400, towards values in the order of 2\% for the eMBB slice, and 0,32\% for the URLLC slice.
\begin{figure}[t!]
\centering
\includegraphics[clip, trim = 0cm 0.2cm 0cm 0cm,width=0.95\columnwidth]{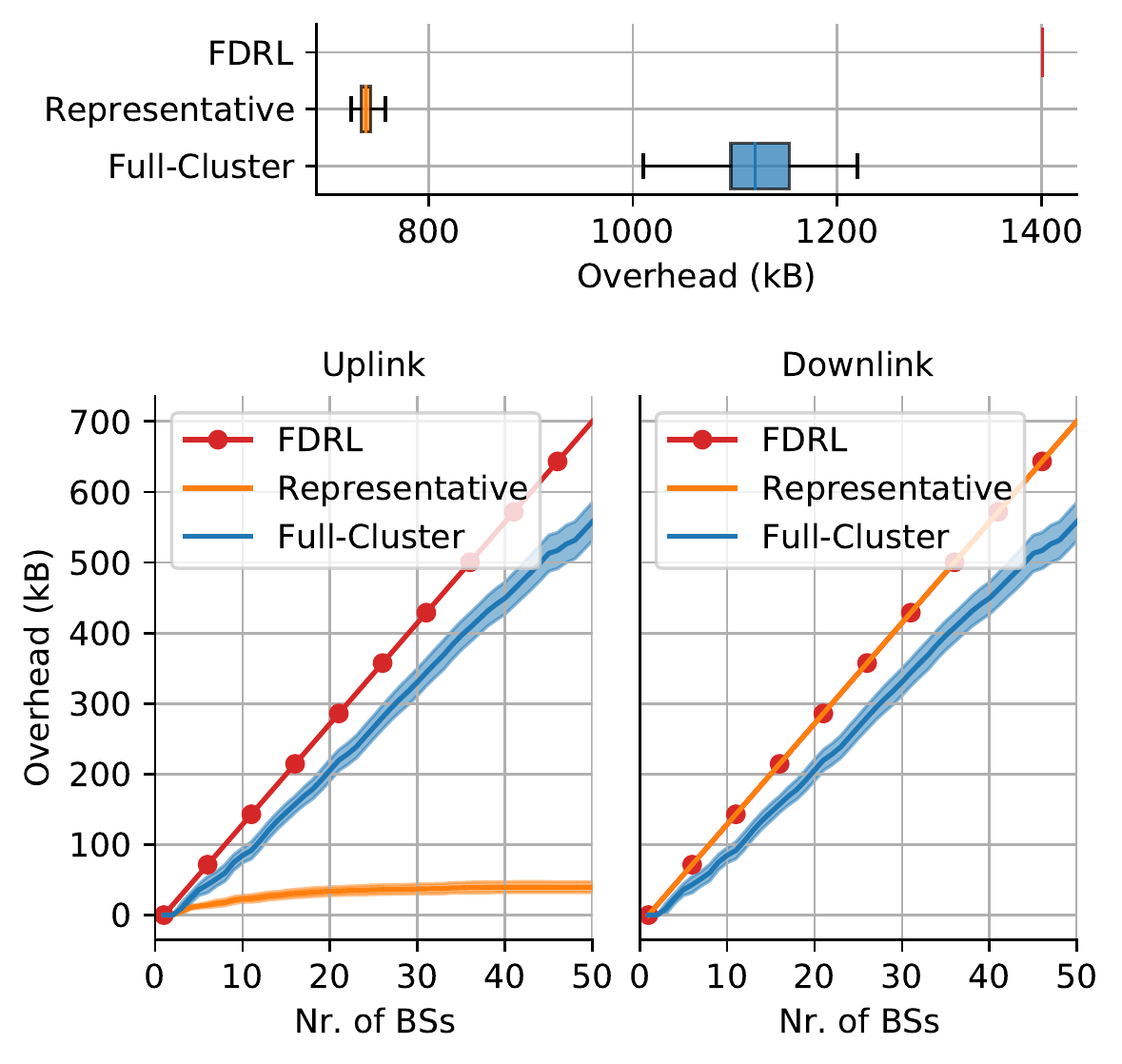}
\caption{\small Communication overhead per federation episode for different federation strategies (top-part) and for different number of BSs deployed (bottom-part). RR and BR federation strategies are referred as Representative.}
\label{fig:Overhead}
\end{figure}
\subsubsection{\change{Communication Overhead Comparison for Different Federation Strategies}}
Federated Learning aims at building global knowledge from the exchange of multiple locally trained models towards a centralized entity.
Such a frequent model exchange however introduces significant communication overhead and synchronization issues, specially in wide scenarios as those considered in our work.
Fig.~\ref{fig:Overhead} compares the model exchange overhead per federation episode resulting from our experiments for a different number of base stations while running the same 3 slices.
In the upper part we focus on the overhead statistical distribution. The benchmark \emph{FDRL} approach assumes the exchange of all the locally trained model weights to derive the federated ones, which implies the highest communication overhead. 
The BR and RR approaches (depicted in the center of the image, and referred to as \emph{Representative}) allow reducing the uplink information exchange by selecting a single representative of each cluster, regardless of the dimensions of the group itself, thus minimizing the communication overhead in each federation episode. It results in less than $800$ kBytes in our settings.
Finally, the \emph{Full-Cluster} approach is characterized by an intermediate average value but higher variance. This is due to the variable size of the DAs clusters, which follows the real-time traffic variations, and the need to exchange the local model weights of each element of the cluster, saving communication resources from those base stations that presented peculiar traffic traces and remain unclustered.

On the lower part of the picture, we differentiate between uplink and downlink model exchange overhead. 
The FDRL approach presents a symmetric behavior matching the model exchange of all the running DAs, in both directions. Conversely, the RR/BR approaches show an asymmetric behavior that favors the upload communication with respect to the downlink one, as only a single DA per cluster shares its local model during the federation process, resulting in a logarithmic trend (with respect to the number of BSs) characterizing the overhead in uplink. This would guarantee better scalability, at the expense of suboptimal performances, as shown in our evaluation.
Finally, the FC approach shows a sublinear trend, with a slower growth rate than the benchmark FDRL, but with significant better performances thanks to the specialization of the DAs. It is safe to assert that the proposed dynamic clustering approach enhances the efficiency of the federated learning scheme, limiting the overall communication overhead with respect to traditional approaches, while providing better performances.
\begin{figure}[h!]
\centering
\includegraphics[clip, trim = 0cm 0.2cm 0cm 0cm,width=0.9\columnwidth]{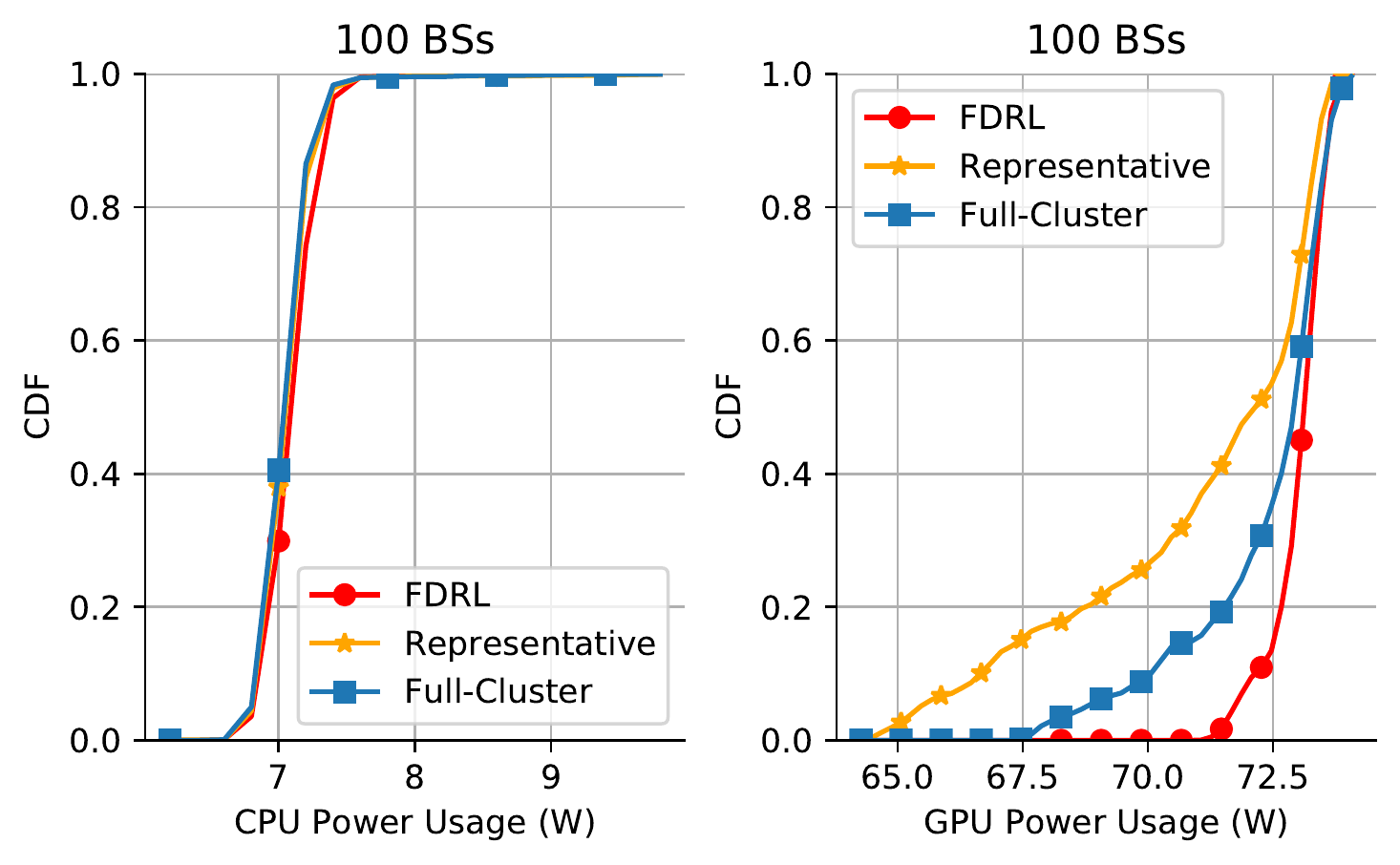}
\caption{\small \change{CPU and GPU power consumption for different federation strategies during agent training.}}
\label{fig:ENERGY}
\end{figure}

\subsubsection{Power Consumption Comparison} 



\change{Energy consumption is an important factor in federeated learning schemes. In Fig.~\ref{fig:ENERGY}, we compare the power consumption of the different DRL strategies during training both in terms of CPU (left-hand side) and GPU (right-hand side), assuming they use the same computational platform as specified at the beginning of Sec.~\ref{sec:perf_eval}.
We use \texttt{nvidia-smi} command line utility\footnote{Part of the NVIDIA management library (NVML). Online available at \url{https://developer.nvidia.com/nvidia-management-library-nvml}}
to retrieve in real-time the energy consumption of the device, whereas for the CPU consumption we monitor the CPU utilization during the training, and consider a proportional fraction of the absorbed power at full computational load as declared by the vendor\footnote {https://ark.intel.com/content/www/us/en/ark/products/192437/intel-xeon-gold-6230-processor-27-5m-cache-2-10-ghz.html}.
As we leverage the GPU hardware to train the models, the different federation strategies exhibit a similar impact on the power consumption of the CPU. Therefore, we focus on the GPU power consumption to better highlight their behavior. The obtained CDFs show that RR/BR schemes present lower consumption compared to FC and FDRL. Besides being positively influenced by the communication overhead variation depicted in Fig.~\ref{fig:Overhead}, such reduced power consumption also results from the limited number of RR/BR agents involved in the federation process (selected through accurate clustering procedures, as shown in Sec.~\ref{subsec:dynamic_clustering}), when compared against baseline approaches.


}

\section{Conclusions}
\label{sec:conclusion}
Major research efforts in the \emph{network slicing orchestration} area focus on designing solutions able to \emph{concurrently and efficiently deal with both spatial and temporal aspects of users' traffic demand}. Due to the distributed nature of the RANs domain, centralized approaches are doomed to provide suboptimal performance and introduce significant communication overhead towards holistic resource controllers.
Tackling such challenging scenario, in this paper we proposed an \emph{FDRL-based architecture for network slice resource orchestration}, where \emph{clusters} of decision agents are dynamically instantiated as virtualized instances with control over base stations radio resources. Enabled by the latest developments in federated learning, our approach allows building specialized knowledge from traffic and mobility patterns by exploiting similarity metrics. Our results show that the proposed  \emph{FDRL-based architecture} poses a trade-off involving the minimization of the communication overhead and the specialization of the decision agents, which in turn affects their accuracy along the resource allocation process.

\comment{
\section*{Acknowledgment}
The research leading to these results has been supported by the H$2020$ MonB5G Project (grant agreement no. $871780$).
}



\vspace{-10mm}

\begin{IEEEbiography}[{\includegraphics[width=1in,height=1.25in,clip,keepaspectratio]{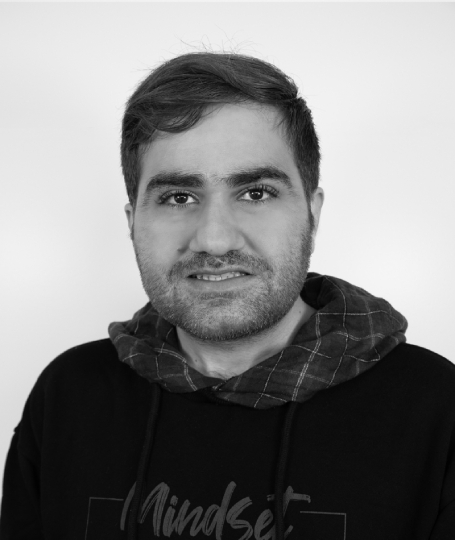}}]{Farhad Rezazadeh}(S'19) is currently a Ph.D. candidate at the Technical University of Catalonia (UPC) and Researcher at CTTC. He is involved in some European H2020 projects and he was awarded first patent connected to H2020 5G-SOLUTIONS project. He was a secondee at NEC Lab Europe and had scientific mission at TUM, TUHH, and UdG. He won multiple European, Government, and IEEE Grants. He also serves/served as Reviewer, Organizing, and TPC member in IEEE. His research interests lie in the area of Lifelong ML, B5G/6G, and Network Slicing.
\end{IEEEbiography}
\vspace{-15mm}
\begin{IEEEbiography}[{\includegraphics[width=1in,height=1.25in,clip,keepaspectratio]{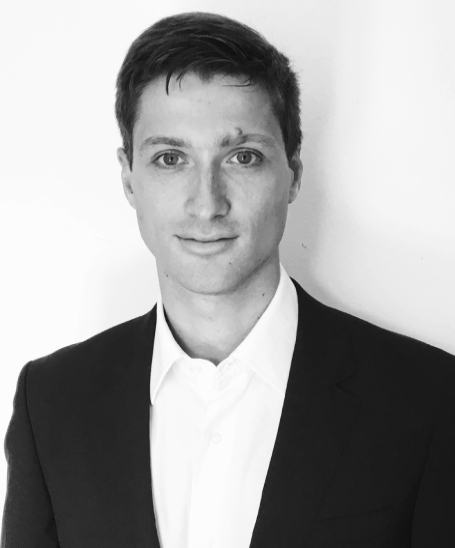}}]{Lanfranco Zanzi}(S'17--M'22) received his B.Sc. and M.Sc. in Telecommunication Engineering from Politecnico of Milan (Italy) in 2014 and 2017, respectively, and the Ph.D. degree from the Technical University of Kaiserlautern (Germany) in 2022. He works as senior research scientist at NEC Laboratories Europe. His research interests include network virtualization, machine learning, blockchain, and their applicability to 5G and 6G mobile networks in the context of network slicing.
\end{IEEEbiography}
\vspace{-15mm}
\begin{IEEEbiography}[{\includegraphics[width=1in,height=1.25in,clip,keepaspectratio]{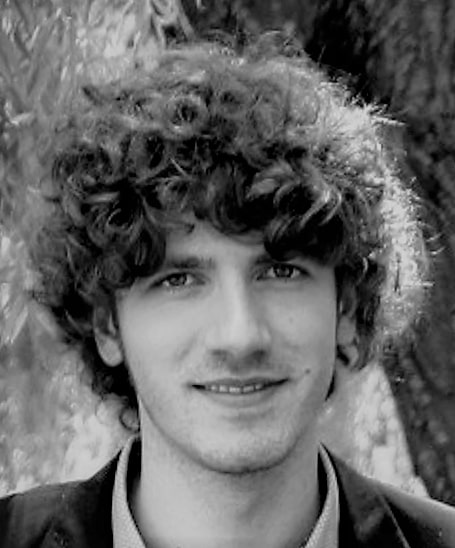}}]{Francesco Devoti} (M'20) received the B.S., and M.S. degrees in Telecommunication Engineering, and the Ph.D. degree in Information Technology from the Politecnico di Milano, in 2013, 2016, and 2020 respectively. He is currently a senior research scientist in the 6G Network group at NEC Laboratories Europe. His research interests include reflective intelligent surfaces, millimeter-wave technologies in 5G and 6G networks, and network slicing.
\end{IEEEbiography}
\vspace{-15mm}
\begin{IEEEbiography}[{\includegraphics[width=1in,height=1.25in,clip,keepaspectratio]{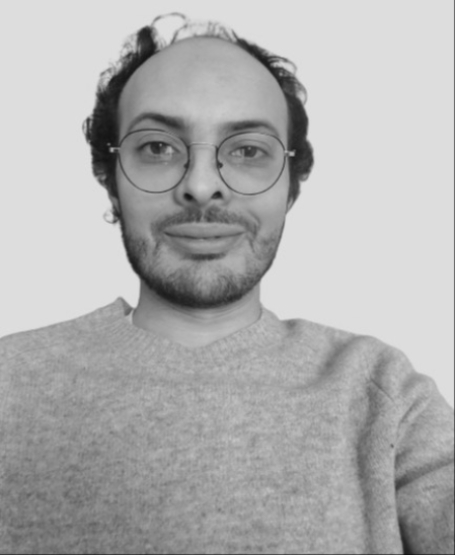}}]{Hatim Chergui} (M'12--SM'22) received the engineering degree in telecommunications from the Institut National des Postes et T\'el\'ecommunications (INPT), Rabat, Morocco, in 2007 and the Ph.D. degree (summa cum laude) in electrical engineering and telecommunications from IMT-Atlantique (T\'el\'ecom-Bretagne), Brest, France, in 2015. He is currently the project manager of the H2020 MonB5G European project and a researcher at CTTC, Spain. He served as a RAN expert at both INWI and Huawei Technologies, Morocco. He was the recipient of the IEEE ComSoc CSIM 2021 Best Journal Paper Award and the IEEE ICC 2020 Best Paper Award. He is an Associate Editor of IEEE Networking Letters.
\end{IEEEbiography}
\vspace{-15mm}
\begin{IEEEbiography}[{\includegraphics[width=1in,height=1.25in,clip,keepaspectratio]{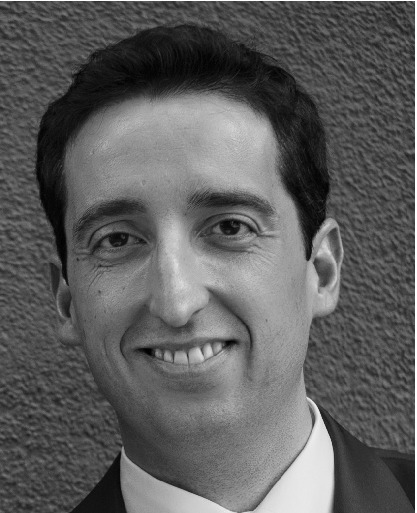}}]{Xavier Costa-P\'erez} (M'06--SM'18) is Head of Beyond 5G Networks R\&D at NEC Laboratories Europe, Scientific Director at the i2Cat R\&D Center and Research Professor at ICREA. His team contributes to products roadmap evolution as well as to European Commission R\&D collaborative projects and received several awards for successful technology transfers. In addition, the team contributes to related standardization bodies: 3GPP, ETSI NFV, ETSI MEC and IETF. Xavier has been a 5GPPP Technology Board member, served on the Program Committee of several conferences (including IEEE Greencom, WCNC, and INFOCOM), published at top research venues and holds several patents. He also serves as Editor of IEEE Transactions on Mobile Computing and Transactions on Communications journals. He received both his M.Sc. and Ph.D. degrees in Telecommunications from the Polytechnic University of Catalonia (UPC) in Barcelona and was the recipient of a national award for his Ph.D. thesis.
\end{IEEEbiography}
\vspace{-125mm}
\begin{IEEEbiography}[{\includegraphics[width=1in,height=1.25in,clip,keepaspectratio]{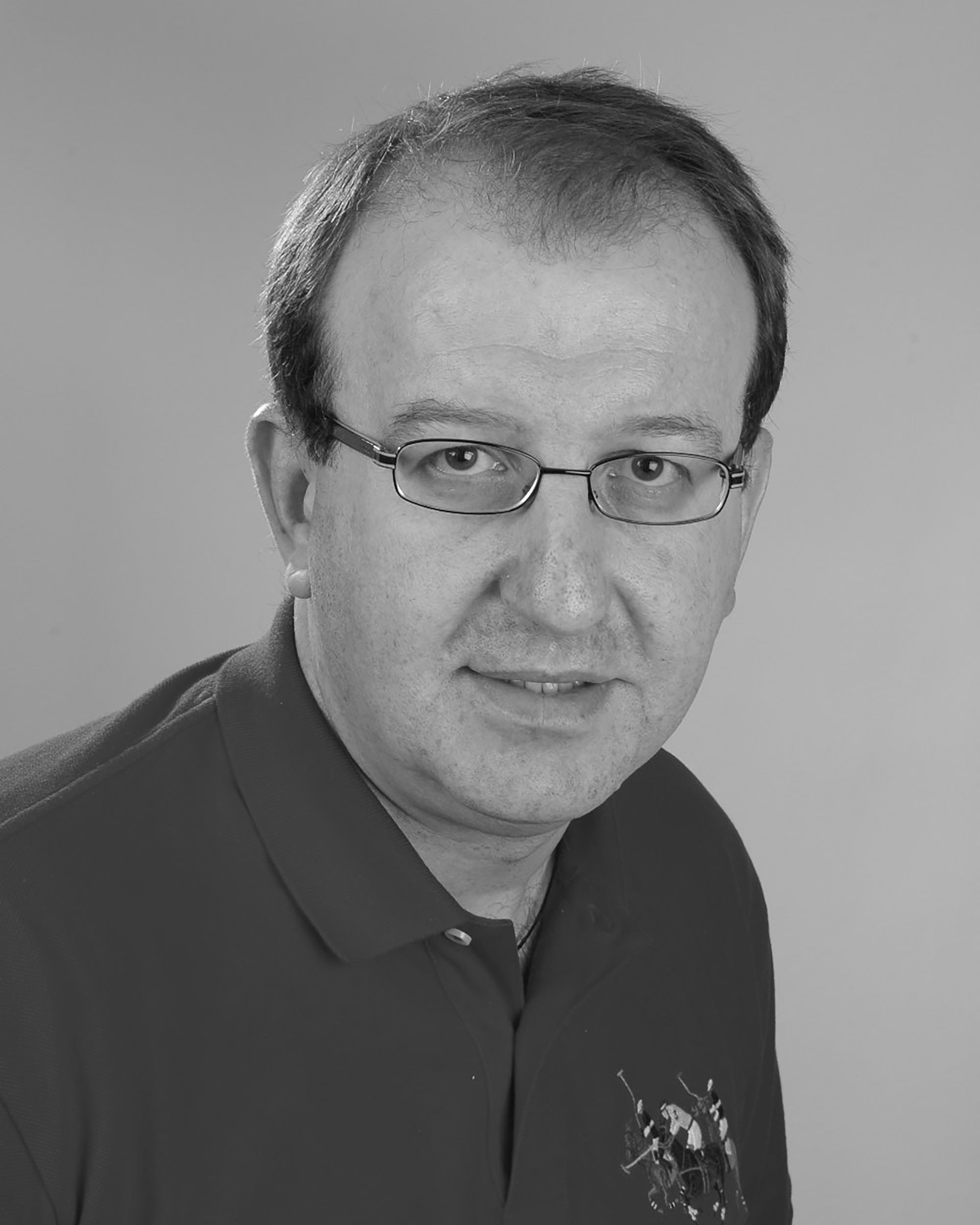}}]{Christos Verekoukis} (SM'07) received the Ph.D. degree from Technical University of Catalonia (UPC), Barcelona, Spain, in 2000. He is currently a scientific director in Iquadrat Informatica, and an Associate Proferssor with the Univerisity of Patras. He has authored 151 journal papers and over 200 conference papers. He is also a coauthor of three books, 14 chapters in other books, and two patents. He has participated in more than 40 competitive projects, and has served as a principal investigator of national projects in Greece and Spain. He is currently the IEEE ComSoc GITC vice-chair and the editor-in-chief of the IEEE Networking Letters.
\end{IEEEbiography}

\end{document}